\documentclass[10pt,aip,cha,twocolumn,amsmath,showpacs,superscriptaddress,reprint]{revtex4-1}
\usepackage{multirow}
\usepackage{rotating}
\usepackage{color}
\def\WT{\widetilde}
\def\lb{\left}
\def\rb{\right}
\def\D{\Delta}
\def\nn{\nonumber}
\def\Piikk{P^{i,i'}_{k,k'}}
\def\Eii{E^{i,i'}}
\def\Pkk{P_{k,k'}}
\def\pik{p^{i}_k}

\begin{document}
\title{Dynamics on Modular Networks with Heterogeneous Correlations}

\author{Sergey Melnik}
\affiliation{MACSI, Department of Mathematics \& Statistics, University of Limerick, Ireland}
\affiliation{Oxford Centre for Industrial and Applied Mathematics, Mathematical Institute, University of Oxford, OX2 6GG, UK}
\affiliation{CABDyN Complexity Centre, University of Oxford, Oxford, OX1 1HP, UK}
\author{Mason A. Porter}
\affiliation{Oxford Centre for Industrial and Applied Mathematics, Mathematical Institute, University of Oxford, OX2 6GG, UK}
\affiliation{CABDyN Complexity Centre, University of Oxford, Oxford, OX1 1HP, UK}
\author{Peter J. Mucha}
\affiliation{Carolina Center for Interdisciplinary Applied Mathematics, Department of Mathematics, University of North Carolina, Chapel Hill, NC 27599-3250, USA}
\affiliation{Institute for Advanced Materials, Nanoscience \& Technology, University of North Carolina, Chapel Hill, NC 27599-3216, USA}
\author{James P. Gleeson}
\affiliation{MACSI, Department of Mathematics \& Statistics, University of Limerick, Ireland}


\begin{abstract}
We develop a new ensemble of modular random graphs in which degree-degree correlations can be different in each module and the inter-module connections are defined by the joint degree-degree distribution of nodes for each pair of modules. We present an analytical approach that allows one to analyze several types of binary dynamics operating on such networks, and we illustrate our approach using bond percolation, site percolation, and the Watts threshold model. The new network ensemble generalizes existing models (e.g., the well-known configuration model and LFR networks) by allowing a heterogeneous distribution of degree-degree correlations across modules, which is important for the consideration of nonidentical interacting networks.
\end{abstract}

\maketitle

\section{Introduction}

It can be very useful to view a network as consisting of a set of heterogeneous, interconnected modules.~\cite{comnotices,Fortunato10} The connections between nodes in the same module or between nodes from different modules tend not to be uniformly random. For example, they might depend on nodes' degrees (or other structural characteristics) or on their module assignments. Furthermore, the meaning of network modules depends significantly on context. For example, a module in a social network might represent a group to which an individual belongs, whereas a partition of a technological network into modules might yield subnetworks that each contain different types of nodes.

Amidst the escalating data deluge, networks that are constructed from multiple interconnected parts or which contain multiple types of nodes have attracted considerable recent interest.~\cite{Kivela13,Vespignani10,Mucha10,Zhou06,Galstyan07,Buldyrev10,Parshani10,Gao12,Brummitt12,Cho10,Leicht09,Allard09,Brummitt10,Mendiola12,Parshani11,Hu11,Tanizawa12} Such networks, for example, can be used to study failures on interdependent power grids, the spread of social influence through multiple media, or transportation via multiple modes of travel.~\cite{Kivela13,Vespignani10} It is therefore important to develop new random-graph models for studying such processes. For example, in order to detect cohesive groups of nodes (i.e., communities) algorithmically in such networks, it is necessary to develop and analyze more sophisticated random-graph null models against which to compare the structure of real networks.~\cite{Mucha10,Bassett13} 

In the present paper, we develop a model of networks that consist of interconnected modules, where each module has its own joint degree-degree distribution, and the connections between nodes from different modules are determined via a joint degree-degree distribution of nodes for each pair of modules. One can think of each module as a separate network, so the aggregate network under consideration constitutes an example of a ``multilayer'' network.~\cite{Kivela13} (Indeed, multi-module networks have many names in the literature --- including ``interdependent networks'', ``coupled networks'', and more.~\cite{Mucha10,Zhou06,Galstyan07,Buldyrev10,Parshani10,Gao12,Brummitt12,Cho10,Leicht09,Allard09,Brummitt10,Mendiola12,Parshani11,Hu11,Tanizawa12}) 

The model that we introduce in this paper (and which we call the \emph{$\Piikk$ network} model) is a generalization of several random-graph models. (Note that we only consider unweighted and undirected networks.) One of them is the well-known configuration model,~\cite{Newman10} in which a degree distribution $p_k$ is specified but connections between stubs (i.e., half-edges) are assigned uniformly at random. Another model, which we call the \emph{$\Pkk$ network} model, describes single-module networks and is defined by a joint degree-degree distribution (i.e., via degree assortativity).~\cite{Newman02} A third model, which we call the \emph{$\Eii$ network} model, generates multi-module networks that have no \textit{a priori} degree-degree correlations.~\cite{Gleeson08a} The so-called LFR (Lancichinetti-Fortunato-Radicchi) synthetic benchmark networks~\cite{Lancichinetti08} for testing community-detection methods constitute a fourth important special case of our model.

We also develop an analytical method that can be used to investigate a broad class of binary-state dynamics operating on networks produced by our generative network model~\cite{Piikk_url} and its special cases above. We employ numerical simulations to test the accuracy of our analytical approach using several synthetic and real-world networks. We show that because our network model incorporates features of network topology that $\Pkk$ and $\Eii$ models are unable to capture, it allows us to better predict the dynamics operating on networks.

The remainder of this paper is organized as follows. In Sec.~\ref{sec2}, we introduce our new $\Piikk$ random-graph ensemble. In Sec.~\ref{sec3}, we present our analytical approach for solving binary-state dynamics on such networks. In Sec.~\ref{sec4}, we show how the same analytical approach can be applied to $\Pkk$ and $\Eii$ networks. In Sec.~\ref{sec5}, we examine several examples of dynamical processes and compare the predictions of our theory with numerical simulations. In Sec.~\ref{sec6}, we use both synthetic and real-world networks to provide examples with more complicated topologies that further justify our new random-graph ensemble. Finally, we present our conclusions in Sec.~\ref{sec7}.


\section{Definition and Construction of $\Piikk$ Networks} \label{sec2}

\begin{figure}[t!]
\centering
\includegraphics[width=0.97\columnwidth]{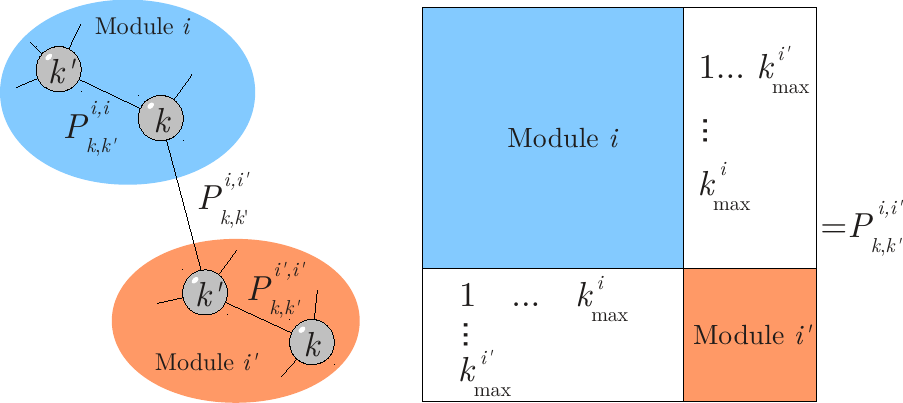}
\caption{Schematic of a network that consists of (left) $M=2$ non-overlapping modules and (right) the corresponding block matrix that encodes the joint degree-degree, module-module distribution $P^{i,i'}_{k,k'}$ (i.e., the probability that a randomly chosen network edge connects a degree-$k$ node from module $i$ to a degree-$k'$ node from module $i'$). The matrix is symmetric and is composed of $M$-by-$M$ blocks. The number of scalar elements in each block is given by the product $k_{\rm max}^i \times k_{\rm max}^{i'}$ of maximum degrees for nodes from the corresponding modules. In essence, the diagonal blocks represent the joint degree-degree distribution for each module. The off-diagonal blocks characterize connections between pairs of modules.}
\label{fig1_Piikk}
\end{figure}

Consider an undirected, unweighted, connected network that consists of $N$ nodes that are distributed across $M$ modules such that each node is a member of exactly one module and $k^{i}_{\rm max}$ is the maximum node degree in module $i$. (We define the node degree to be the total number of a node's neighbors across all modules. It can, of course, also be desirable to be more nuanced and consider multiple types of degrees.~\cite{Kivela13}) We define $\Piikk$ to be the probability that a randomly chosen edge connects a degree-$k$ node from module $i$ to a degree-$k'$ node from module $i'$.~\footnote{We define $\Piikk$ rigorously as follows: Choose a network edge uniformly at random, and label its end nodes (also uniformly at random) as $A$ and $B$. It follows that $\Piikk$ is the joint probability that $A$ is a degree-$k$ node in module $i$ and $B$ is a degree-$k'$ node in module $i'$. Therefore, $(2-\delta_{i,i'}\delta_{k,k'})\Piikk$ is the probability that a randomly chosen edge connects a degree-$k$ node in module $i$ and a degree-$k'$ node in module $i'$ (because, in this case, one no longer distinguishes between the nodes at the two ends of the edge).} 
In the context of social networks, $\Piikk$ thereby allows modules (e.g., social circles) with different levels of degree homophily. As we illustrate in Fig.~\ref{fig1_Piikk}, one can think of the tensor $[\Piikk]$ with elements $\Piikk$ as an $M$-by-$M$ block matrix. Each block ($i,i'$) is a $k^{i}_{\rm max}$-by-$k^{i'}_{\rm max}$ matrix of scalars, so $[\Piikk]$ contains $\lb(\sum_{i=1}^{M} k_{\rm max}^i\rb)^2$ scalar entries. Each block on the main diagonal is given (up to a proper normalization) by the joint degree-degree distribution for that module, and off-diagonal blocks represent connections between pairs of different modules.

To construct~\cite{Piikk_url} an $N$-node network drawn from the ensemble of random graphs with specified $\Piikk$ distribution, we first need to calculate how many edges of each type are in the network. We therefore calculate the degree distribution 
\begin{equation}
    p_k = \lb(\sum_{i,i',k'} \frac{\Piikk}{k}\rb) \lb/ \lb(\sum_{i,i',k,k'} \frac{\Piikk}{k}\rb)\rb.,
\end{equation}	
and we note that the mean degree is $z = \sum_k k p_k$. Consequently, the number of edges of type $\{(i,k)$,$(i',k')\}$ (i.e., edges that connect a degree-$k$ node from module $i$ to a degree-$k'$ node from module $i'$) is $\frac12 z N \Piikk$. We then create the required number of edges of each type and place degree-$k$ nodes in module $i$ by collecting $k$ edge ends of type $(i,k)$ uniformly at random. When we have gathered all edge ends into such bunches, we have constructed the desired network.

Importantly, one can obtain a $\Piikk$ network from a real-world network (or any other network) by rewiring its edges using an algorithm that preserves the $\Piikk$ distribution but otherwise randomizes connections between nodes.~\footnote{To do this, one can employ the following network rewiring algorithm: Choose an edge of the network uniformly at random. Denote its end nodes by $A$ and $B$, their corresponding modules by $i_A$ and $i_B$, and their degrees by $k_A$ and $k_B$. Choose another edge uniformly at random from the set of edges that have one end-node of degree $k_A$ in module $i_A$. This edge connects nodes $C$ and $D$ from modules $i_A$ and $i_D$, whose respective node degrees are $k_A$ and $k_D$. One now rewires the two chosen edges to obtain the edges $AD$ and $CB$ instead of $AB$ and $CD$. In applying this algorithm, we also take care to avoid creating multiple-edges and self-edges. This rewiring scheme does not affect the degrees or modules of the rewired nodes, but it randomizes connections between them. Applying this scheme repeatedly significantly reduces the density of triangles and thereby reduces the local clustering.}

\section{Analytical Calculation of Dynamics on $\Piikk$ Networks} \label{sec3}

In this section, we consider binary-state dynamics on networks --- in which nodes can be in one of two possible states, which we call ``inactive'' ($0$) and ``active'' ($1$) --- and present analytical expressions for calculating the fraction of active nodes in a network starting with some (possibly zero) fraction of initially active nodes. Even though binary dynamics are the simplest type of dynamical process that can occur on a network, they have repeatedly been very insightful on a large variety of (much more complicated) real systems.~\cite{Newman03a,Newman10,Barrat08,Gleeson13,Watts02,Grassberger83,Centola07a,Centola07b} In a binary-dynamics model, an idea is either accepted or not accepted, a purchase is either made or not made, an individual is either infected with a disease or is not infected, a system component has either failed or it has not, a profile picture is changed or it is not,~\cite{FB-equal,FB-equal2} and so on. On an appropriate timescale, it can be very insightful to use binary dynamics as a simple model for very complicated situations even when the dynamics is not binary on a longer timescale.

We consider a broad class of binary dynamics that can be described in terms of ``response functions" $F_i(m,k)$, which describe local interactions among neighboring nodes and which can be defined independently for each module $i$. The response function $F_i(m,k)$ gives the probability that a degree-$k$ node in module $i$ becomes active when it has $m$ active neighbors. The analytical method that we employ is based on pairwise interactions between nodes,~\cite{Gleeson07a,Gleeson08a,Gleeson13} and it requires that the response functions $F_i(m,k)$ be non-decreasing functions of $m$ for any fixed $k$. This condition reflects the effect of positive externalities: a node is more likely to become active when it has more active neighbors. Examples of dynamics that satisfy this requirement include bond percolation, site percolation, the Watts threshold model, and the calculation of $k$-core sizes.~\cite{Gleeson08a} In Sec.~\ref{sec5}, we will discuss the first three of these examples.

To calculate the fraction of active nodes at discrete time $n$, we employ the approach introduced in Refs.~\onlinecite{Gleeson07a,Gleeson08a} and which has proven to be very fruitful.~\cite{Melnik13,Gleeson13,Gleeson09a,Gleeson09b,Hackett11,Gleeson10,Gleeson08b,Brummitt12,Yagan12,Payne11,Ikeda10} We define $q^i_k(n)$ to be the probability that a degree-$k$ node in module $i$ is active at time step $n$, given that at least one of its neighbors is inactive. The probability that a neighbor of an inactive degree-$k$ node in module $i$ (at time step $n$) is active is then given by the expression
\begin{align}\label{qikbar}
    \bar q^i_k(n) = \frac{\sum_{i'} \sum_{k'} P^{i,i'}_{k,k'} q^{i'}_{k'}(n)}{\sum_{i'}\sum_{k'} P^{i,i'}_{k,k'}}\,,
\end{align}
because a degree-$k$ node of module $i$ has a neighbor of degree $k'$ node in module $i'$ with probability
\begin{equation*}
    \frac{\Piikk}{\sum_{i'}\sum_{k'}\Piikk}\,.
\end{equation*}	
Therefore, starting with the initial condition $q^i_k(0)=\rho^i_k(0)$, where $\rho^i_k(0)$ is the initially active fraction of degree-$k$ nodes in module $i$, one can compute the values of $q^i_k(n)$ using the recurrence equation
\begin{align}\label{qikn}
    q^i_k&(n+1)= \rho^i_k(0) + \lb(1-\rho^i_k(0)\rb) \\
\nn &\times \sum_{m=0}^{k-1} \binom{k-1}{m} \lb(\bar q^i_k(n) \rb)^m \lb(1-\bar q^i_k(n)\rb)^{k-1-m} F_i(m,k)\,.
\end{align}
Equation~\eqref{qikn} describes a situation in which a node that was initially inactive becomes active when it has a sufficient number of active neighbors (given that one of its neighbors is inactive).

The probability that a degree-$k$ node in module $i$ is active at time step $n+1$ (i.e., the fraction of active degree-$k$ nodes in module $i$) is calculated from
\begin{align}\label{rhoikn}
    \rho^i_k(n+1)&= \rho^i_k(0) + \lb(1-\rho^i_k(0)\rb)\\
\nn &\times \sum_{m=0}^k \binom{k}{m} \lb(\bar q^i_k(n) \rb)^m \lb(1-\bar q^i_k(n)\rb)^{k-m} F_i(m,k)\,.
\end{align}
One then calculates that the fraction of active nodes in module $i$ at time step $n$ is
\begin{align}
    \rho^i(n)= \sum_k \pik \rho^i_k(n)\,,
\end{align}
where 
\begin{equation}\label{pik}
    \pik = \lb(\sum_{i',k'}\frac{1}{k}\Piikk\rb)\lb/\lb(\sum_{i',k',k}\frac{1}{k}\Piikk\rb),\rb.
\end{equation}

is the degree distribution for nodes in module $i$, and the aggregate active fraction of nodes in the entire network is
\begin{align}
    \rho(n)= \sum_i \frac{N_i}{N} \rho^i(n)\,,
\end{align}
where 
\begin{equation}
    N_i = N \lb(\sum_{k,k',i'} \frac1k \Piikk\rb)\lb/\lb(\sum_{i,i',k,k'} \frac1k \Piikk\rb)\rb.
\end{equation}	
is the number of nodes in module $i$.


\section{Application to $\Pkk$ and $\Eii$ networks} \label{sec4}

The $\Piikk$ random-graph ensemble generalizes two well-known models: $\Pkk$ networks, which consist of a single module defined by a joint degree-degree distribution of nodes $\Pkk$,~\cite{Newman02,Newman03c} and the uncorrelated multi-module network ensemble (so-called $\Eii$ networks), which are defined by the degree distribution $\pik$ of each module and the probability $\Eii$ that a given network edge connects a node from module $i$ to a node from module $i'$.~\cite{Newman03c} The quantities that define these networks can be calculated from the $\Piikk$ distribution using the formulas
\begin{equation}\label{relations}
  \Pkk = \sum_{i,i'} \Piikk\,, \qquad
  \Eii = \sum_{k,k'} \Piikk\,, 
\end{equation}
and $\pik$ is obtained from Eq.~\eqref{pik}.

Equations~\eqref{qikbar}--\eqref{rhoikn} apply to both $\Pkk$ networks and $\Eii$ networks. Consequently, one can derive results for $\Pkk$, $\Eii$, and $\Piikk$ networks by using only an implementation for $\Piikk$ networks. For example, to obtain results for $\Eii$ networks using Eqs.~\eqref{qikbar}--\eqref{rhoikn}, one needs to input the corresponding \emph{uncorrelated} $\WT \Piikk$ matrix that is constructed from a given mixing matrix $\Eii$ and degree distribution $p^i_k$ of nodes in module $i$ using the formula
\begin{align}
    \WT P^{i,i'}_{k,k'} = \Eii \frac{ k k' \pik p^{i'}_{k'} }{\sum_k k \pik \sum_{k'} k' p^{i'}_{k'}}\,.
\end{align}
To obtain results for $\Pkk$ networks, one can directly use the joint degree-degree distribution matrix $\Pkk$ in Eqs.~\eqref{qikbar}--\eqref{rhoikn} because it represents a single-module $\Piikk$ network. When considering example networks, we use this technique to compare the results given by Eqs.~\eqref{qikbar}--\eqref{rhoikn} for $\Piikk$, $\Pkk$, and $\Eii$ networks (and we refer to the corresponding results as $\Piikk$, $\Pkk$, and $\Eii$ theories).

In general, we expect $\Piikk$ theory to have a better match than $\Pkk$ and $\Eii$ theories to the results of numerical simulations. This is the case because $\Piikk$ theory provides more information about network structure.  For example, unlike $\Pkk$ theory, our new $\Piikk$ theory is able to distinguish between nodes of the same degree that belong to different modules. Unlike $\Eii$ theory, our new theory is able to capture degree-degree correlations that exist both within and between modules.


\section{Examples of Dynamics on Modular Synthetic $\Eii$ Networks} \label{sec5}

In this section, we consider several examples of dynamical processes on modular networks. We compare computational results to theoretical predictions from Eqs.~\eqref{qikbar}--\eqref{rhoikn}. We use $\Eii$ networks, which we showed in Sec.~\ref{sec4} are a special case of $\Piikk$ networks. We consider examples with more complicated specifications in Sec.~\ref{sec6}.

In $\Eii$ networks, each node is a member of exactly one of a network's $M$ modules. The nodes draw their degrees $k$ from the degree distribution $p^i_k$, which is specified separately for each module $i$. Each edge from a node belonging to module $i$ is connected to a node from module $j$ with a fixed probability. These probabilities can be described conveniently by an $M$-by-$M$ mixing matrix $\Eii$, whose representative element $e_{ij}$ gives the probability that a randomly chosen edge in a network connects a node from module $i$ to a node from module $j$. There are $N_i$ nodes in module $i$, and the network is otherwise random.

\begin{figure}[ht]
\centering
\includegraphics[width=0.8\columnwidth]{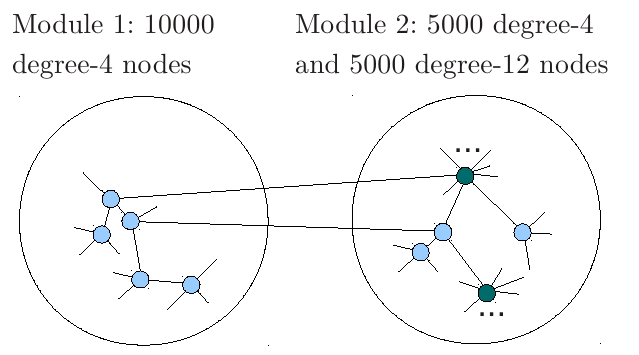}
\caption{Schematic of the $\Eii$ network described in the text. Module 1 consists only of degree-4 nodes, and module 2 consists of nodes of degrees 4 and 12 in proportion 1:1. (The dark nodes have degree 12, and the light ones have degree 4.)}
\label{fig2_Network_sketch_Eii}
\end{figure}

In the examples below, we use an $\Eii$ network that we designed to demonstrate the advantage of $\Piikk$ theory over $\Pkk$ theory. The network consists of two modules with $N_i$=10000 nodes each. The first module contains only nodes of degree 4. The second module contains nodes of degrees 4 and 12 in proportion 1:1, and the mixing matrix
\begin{align}
\Eii =\frac{1}{1200}\lb( 
  \begin{array}{cc}
      399 & 1\\
      1 & 799
  \end{array}\rb)\label{Eii}\,
\end{align}
defines the interconnections between the modules. Such a network is thus an example of a $(k_1,k_2)$-regular graph~\cite{Melnik13} (i.e., a graph in which all nodes have either degree $k_1$ or degree $k_2$). In Fig.~\ref{fig2_Network_sketch_Eii}, we show a schematic of a network drawn from this ensemble. In the following subsections, we investigate several dynamical processes on such networks.

\subsection{Bond Percolation}

We begin by considering bond percolation, which has been studied extensively on networks.~\cite{Newman10} In bond percolation, network edges are deleted (or labeled as ``unoccupied'') with probability $1-p$, where $p$ is called the bond occupation probability. One can measure the effect of bond deletions on aggregate graph connectivity in the limit of infinitely many nodes using the fractional size of the giant connected component (GCC) at a given value of $p$. (In this paper, we use the terminology GCC for finite graphs as well; one can alternatively use the term ``largest connected component'' for finite graphs.)  The fractional size of the GCC is the number of nodes in the GCC of a network divided by the number of nodes in the network. Bond percolation has been used as a simple model for biological epidemics.~\cite{Grassberger83,Newman03a,Newman10} In such a context, $p$ is related to the mean transmissibility of a disease, so the GCC is used to represent the size of an epidemic outbreak (and to give the steady-state infected fraction in a susceptible-infected-recovered model).

To apply Eqs.~\eqref{qikbar}--\eqref{rhoikn} to bond percolation, observe that network edges are occupied with probability $p$ and that nodes become infected (i.e., active) if they are connected to an infected node by an occupied edge. Therefore, a node with $m$ active neighbors has a probability of $(1-p)^m$ of not becoming infected, so the response function is
\begin{align} \label{F_bperc}
    F_i(m,k) = 1-(1-p_i)^m\,.
\end{align}
If different modules $i$ have different values of $p_i$, one obtains what is known as ``semidirected bond percolation'':~\cite{Allard09} every inter-module edge of the original undirected network is replaced by two directed edges that run in opposite directions, and edges that point to nodes of module $i$ are occupied with probability $p_i$. For simplicity, however, we consider classical bond percolation, so we assume that $p_i=p$ for all modules $i$.

\begin{figure}[ht]
\centering
\includegraphics[width=0.97\columnwidth]{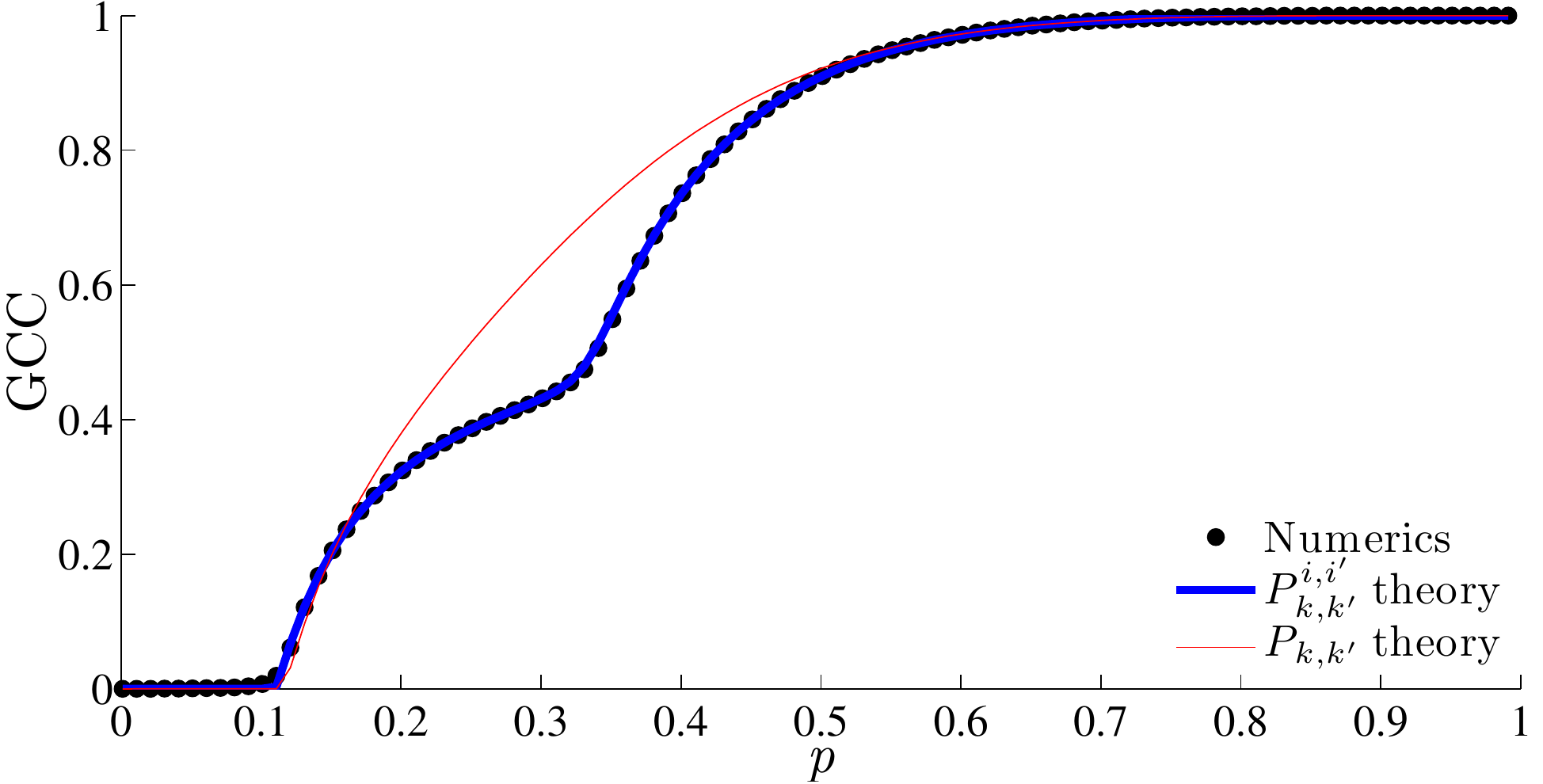}
\caption{GCC size versus bond occupation probability for a $\Eii$ network that consists of two modules that each have $N_i=10000$ nodes. The first module contains only degree-4 nodes. The second module consists of nodes of degrees 4 and 12 in proportion 1:1 (see Fig.~\ref{fig2_Network_sketch_Eii}). The mixing matrix $\Eii$ is given by Eq.~\eqref{Eii}. The result from $P^{i,i'}_{k,k'}$ theory is indistinguishable from numerical simulations, whereas $P_{k,k'}$ theory fails to describe the observed result.}
\label{fig3_bperc_Eii}
\end{figure}

Given a network adjacency matrix, we calculate the distributions $\Piikk$ and $\Pkk$ and then use them and the response function~\eqref{F_bperc} in Eqs.~\eqref{qikbar}--\eqref{rhoikn} to predict the GCC size for a particular value of $p$. (One can also obtain an analytical result for GCC sizes on $\Pkk$ networks using Eq.~(12) of Ref.~\onlinecite{Vazquez03}.) In Fig.~\ref{fig3_bperc_Eii}, we compare the predictions from $\Piikk$ and $\Pkk$ theories.

We perform numerical calculations of the GCC size by applying the algorithm of Ref.~\onlinecite{Newman01e} to the network adjacency matrices and display the results as black disks in Fig.~\ref{fig3_bperc_Eii}. It is apparent from Fig.~\ref{fig3_bperc_Eii} that $\Piikk$ theory outperforms $\Pkk$ theory on this example. The reason for the poor performance of $\Pkk$ theory is as follows. The degree-4 nodes in module 1 have a much smaller probability to have degree-12 nodes as neighbors than do the degree-4 nodes in module 2 (which, in turn, affects the probability of belonging to a GCC). As one can see in the figure, $\Pkk$ theory fails to capture these differences, as it deals only with a single type (the `average') of degree-4 nodes in the network. By contrast, $\Piikk$ theory is able to distinguish between degree-4 nodes in module 1 and degree-4 nodes in module 2. It thus yields the correct prediction in Fig.~\ref{fig3_bperc_Eii}.

To understand the peculiar shape of the numerical curve in Fig.~\ref{fig3_bperc_Eii}, recall that the bond-percolation threshold (i.e., the value of $p$ at which a GCC appears as $p$ increases) for random (configuration-model) networks in which all nodes have degree $k$ is given by $p_{th}=1/(k-1)$.~\cite{Newman10} In Fig.~\ref{fig3_bperc_Eii}, the percolation threshold is dominated by degree-12 nodes from module 2, but its value is slightly shifted to the right from $1/11$ because of the presence of degree-4 nodes in the same module. The step that appears at $p=1/3$ is due to percolation of module 1, which consists entirely of degree-4 nodes. More generally, this also illustrates that $(k_1,k_2)$-regular graphs can be very useful for probing the behavior of dynamics on networks.~\cite{Melnik13}


\subsection{Site Percolation}

We now consider site percolation, in which we remove nodes (along with the edges connected to those nodes) instead of removing edges.~\cite{Newman10} Site percolation can be used as a toy model of the vaccination of individuals against a disease. In the context of a disease that spreads through a network of contacts, vaccinated individuals do not contribute to the spread of the disease and can be construed as having been removed from a network. In this case, the GCC again represents the size of an epidemic outbreak.

To model the site-percolation process, nodes are occupied with a selected probability and occupied nodes become active (i.e., infected) if they have one or more active neighbors. Unoccupied nodes can never become active. The response function for site percolation is thus
\begin{align}
  F_i(m,k) =\lb\{\begin{array}{cl} 0\,,& {\rm if} \quad m=0 \\
                              Q^i_k\,, & {\rm otherwise} \end{array}\rb. \,,
\end{align}
where $Q^i_k$ is the occupation probability of degree-$k$ nodes in module $i$.

\begin{figure}[ht]
\centering
\includegraphics[width=0.97\columnwidth]{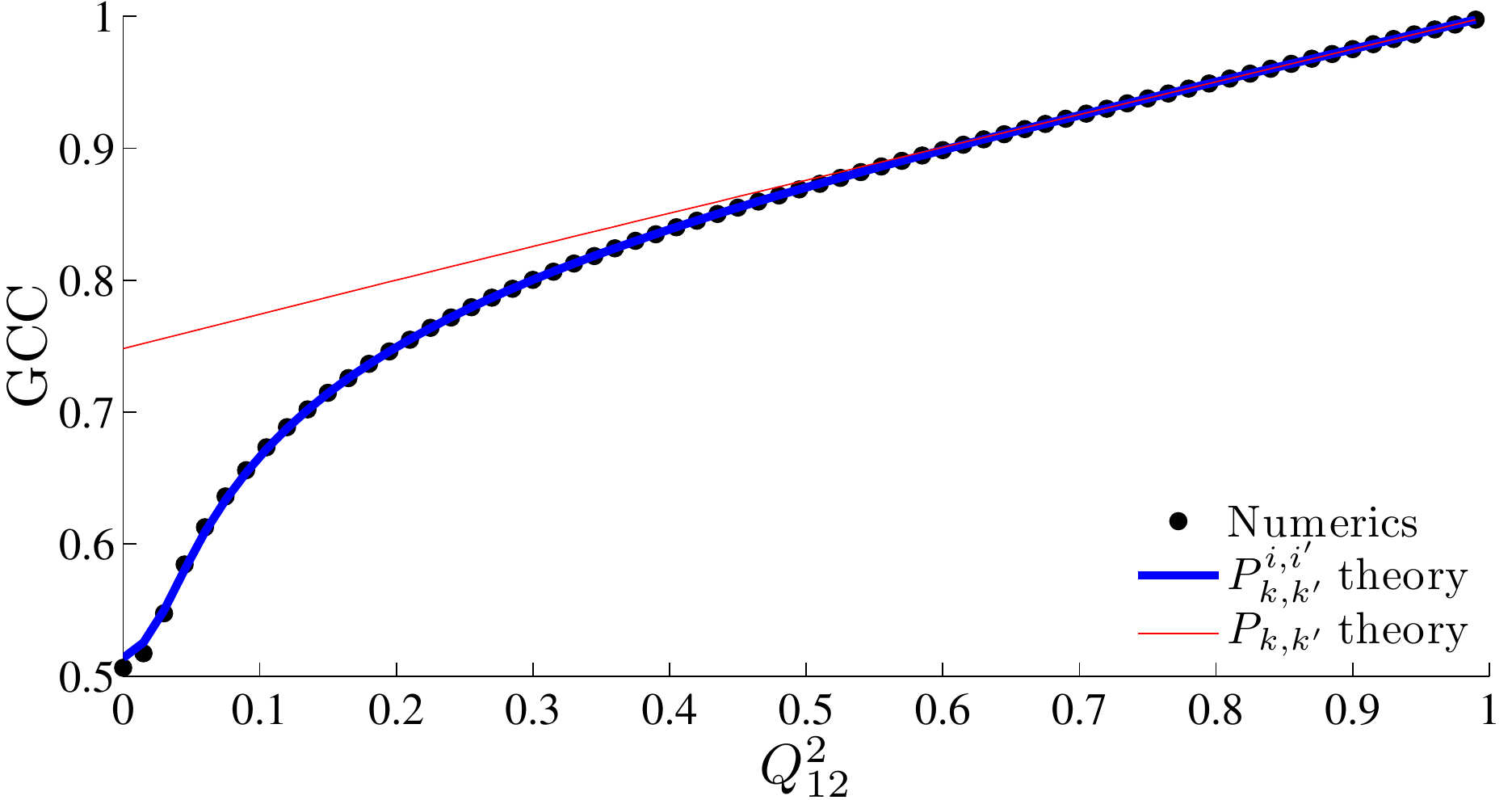}
\caption{Site percolation on the same $\Eii$ network as in Fig.~\ref{fig3_bperc_Eii}. We plot the size of the GCC versus the occupation probability of degree-12 nodes.}
\label{fig4_sperc_Eii}
\end{figure}

In the example in Fig.~\ref{fig4_sperc_Eii}, we compare numerically calculated GCC sizes for the site-percolation process with those obtained using $\Piikk$ and $\Pkk$ theories. We consider the same two-module network as in Fig.~\ref{fig3_bperc_Eii}. Recall that the first module contains only degree-4 nodes, whereas the second module is a mixture of degree-4 and degree-12 nodes. For simplicity and in order to be able to compare results from $\Piikk$ and $\Pkk$ theories directly, we fix all $Q^i_4=1$ and calculate the GCC size as we vary $Q^2_{12}$, which we recall is the occupation probability of degree-12 nodes in module $2$. (In other words, we have constructed this example so that degree-12 nodes are occupied with probability $Q^2_{12}$ and all other network nodes are occupied with probability $1$.) The results of direct numerical computation are approximated very well by the results of $\Piikk$ theory but not by those of $\Pkk$ theory.


\subsection{Watts Threshold Model} 

About a decade ago, Watts introduced a simple model for the spread of cultural fads.~\cite{Watts02} It allows one to examine how a small initial fraction of early adopters can lead to a global cascade of adoption via a social network and it distinguishes between ``simple'' and ``complex'' contagions.~\cite{Centola07a,Centola07b} In the Watts model, each node of a network is randomly assigned a fixed threshold $R$ from a specified probability distribution. A degree-$k$ node becomes active if at least a threshold fraction $R$ of its $k$ neighbors are active (i.e., if $m/k\ge R$). Cascades of activations can be initiated by randomly activating a \emph{seed fraction} $\rho(0)$ of the nodes or, in a more general setting, by randomly activating a fraction $\rho_k^i(0)$ of degree-$k$ nodes in module $i$. The response function for the Watts threshold model is
\begin{align}
    F_i(m,k) = C_i(m/k)\,,
\end{align}
where $C_i$ is the cumulative distribution function of nodes' thresholds in module $i$. If, for example, all nodes in module $i$ have the same threshold $R_i$, then the response function becomes
\begin{align} \label{F_Watts}
    F_i(m,k) =\lb\{\begin{array}{cl} 1\,,& {\rm if} \quad m/k\ge R_i \\
                                     0\,,& {\rm otherwise} \end{array}\rb. \,.
\end{align}

\begin{figure}[ht]
\centering
\includegraphics[width=0.97\columnwidth]{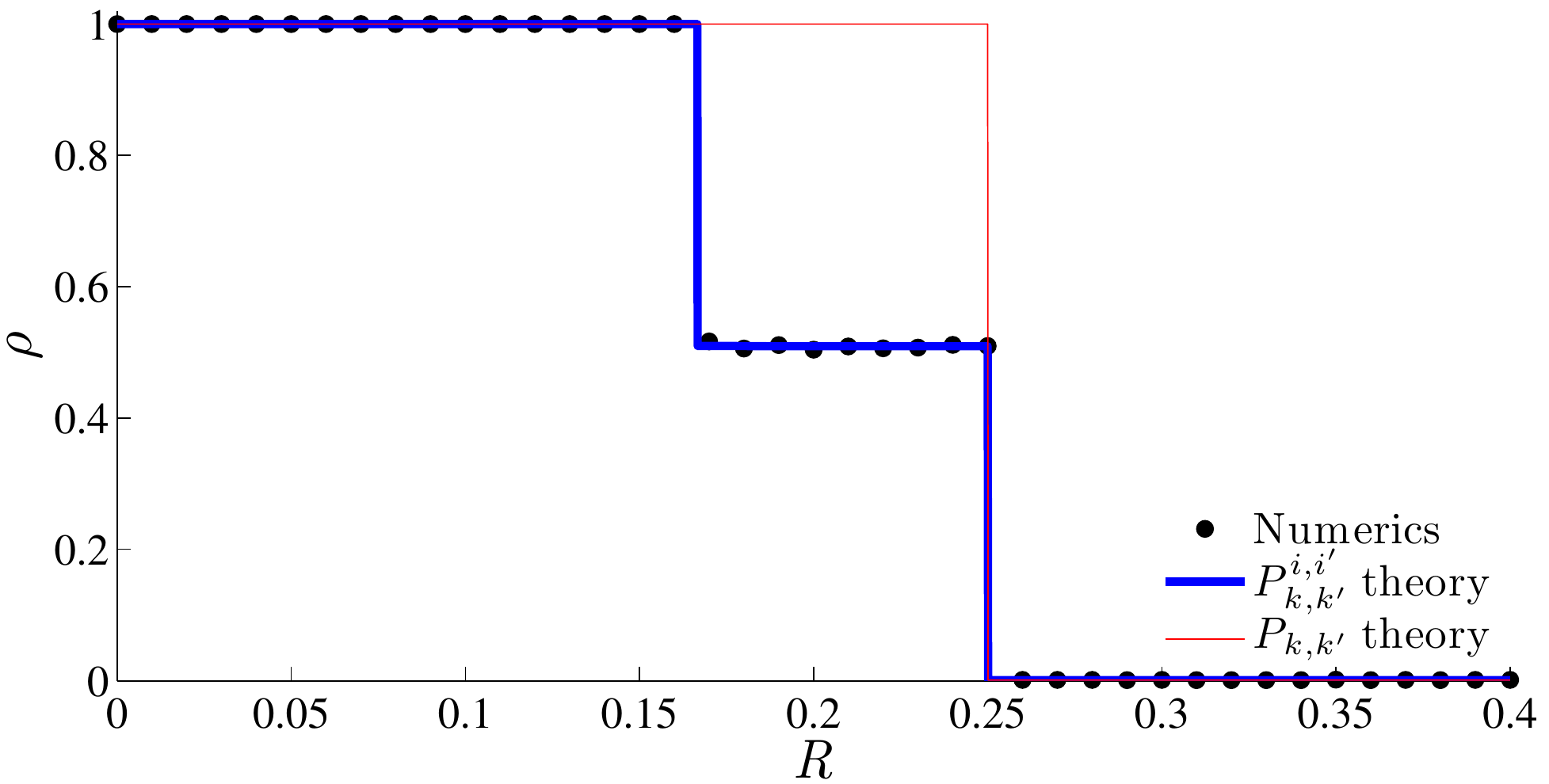}
\caption{Watts threshold model on the same $\Eii$ network as in Fig.~\ref{fig3_bperc_Eii}. We plot the final fraction of active nodes versus the threshold $R$, which is identical for all nodes. Initially, 0.1\% of the nodes (chosen uniformly at random) are active. 
}
\label{fig5_Watts_vs_R_Eii}
\end{figure}

In Fig.~\ref{fig5_Watts_vs_R_Eii}, we show the final fraction of active nodes for the Watts threshold model versus the threshold $R$, which for simplicity we take to be the same for all nodes. In this figure, we use the same $\Eii$ network as for Figs.~\ref{fig3_bperc_Eii} and~\ref{fig4_sperc_Eii}. Initially, we activate $0.1\%$ of nodes, which we choose uniformly at random from the entire network. Nodes subsequently become active according to Eq.~\eqref{F_Watts}. The results of direct numerical simulations are captured accurately by $\Piikk$ theory but not by $\Pkk$ theory,~\cite{Gleeson08a} highlighting the fact that modular structure is important for this process.

In Fig.~\ref{fig5_Watts_vs_R_Eii}, the curve from $\Piikk$ theory exhibits sharp transitions at $R=1/4$ and $R=1/6$. The transition at $R=1/4$ is due to degree-4 nodes in module 1 becoming active when they have at least one active neighbor, so all nodes in module 1 are active at the end of the process. However, module 2 has not experienced an activation cascade at this value of $R$ because most edges in it connect a pair of degree-12 nodes, which collectively remain inactive. The transition at $R=1/6$ occurs when degree-12 nodes become active as a result of having only two active neighbors. In this situation, both modules eventually experience activation cascades. However, there is only one transition in the curve from $\Pkk$ theory. This occurs at $R=1/4$, which is the maximum threshold value for which having a single active neighbor is sufficient to activate a degree-4 node. The curve has the given shape because $\Pkk$ theory mixes all degree-4 and degree-12 nodes in a single module. Therefore, degree-12 nodes are surrounded by many more (active) degree-4 nodes than in module 2 in the actual network. This leads to an erroneous prediction of an activation cascade among degree-12 nodes.


\section{Advanced Examples} \label{sec6}

\subsection{Synthetic $\Piikk$ Networks}

\begin{figure}[ht]
\centering
\includegraphics[width=0.8\columnwidth]{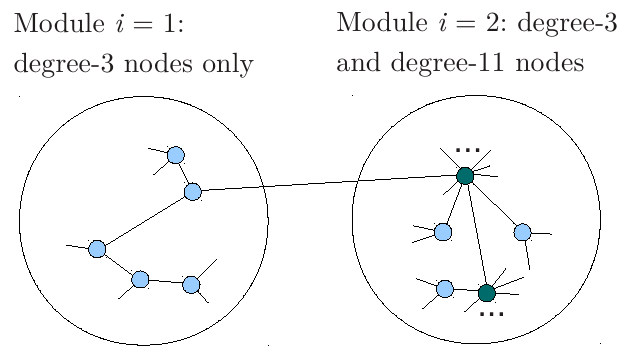}
\caption{Schematic of a $\Piikk$ network with the $\Piikk$ matrix~\eqref{Piikk_mat}. Module 1 contains only degree-3 nodes, and module 2 contains nodes of degree 11 (dark disks) and of degree 3 (light disks).}
\label{fig6_Network_sketch_Piikk}
\end{figure}

The examples that we illustrated in Figs.~\ref{fig3_bperc_Eii},~\ref{fig4_sperc_Eii}, and~\ref{fig5_Watts_vs_R_Eii} can be captured using $\Eii$ theory, because the synthetic network that we used in those figures is drawn from an $\Eii$ ensemble. To demonstrate the full advantage of the $\Piikk$ network model and theory, we generate synthetic $\Piikk$ networks that cannot be constructed using the simpler $\Eii$ network model and investigate dynamical systems on such networks. As an example, consider an ensemble of graphs generated according to the following $\Piikk$ matrix:\footnote{This is a compact form of a matrix of the type that we showed in Fig.~\ref{fig1_Piikk}.  We omit rows and columns that contain only $0$ entries.}
\begin{align}
    \begin{tabular}{c@{}c@{}c@{\quad}c@{\quad}ccllc}
        && $i=1$& \multicolumn{2}{c}{$i=2$} & &&&\\ 
        && $k=3$ & $k=3$& $k=11$ &\vspace{10pt} \\ 
        &\multirow{3}{*}{$\lb.\begin{array}{c}\\ \\ \\ \end{array}\rb($}& 20& 0 & 1 &\multirow{3}{*}{$\lb)\begin{array}{c}\\ \\ \\ \end{array}\rb.$}&{$k'=3$}& $i'=1$  \\
        $\Piikk = \displaystyle\frac{1}{41}$&& 0 & 0 & 9 && $k'=3$ & \multirow{2}{*}{$i'=2$} \\ 
        && 1 & 9 & 1 &&{$k'=11$}\\ 
    \end{tabular}
\label{Piikk_mat}
\end{align}
Networks defined by the $\Piikk$ matrix~\eqref{Piikk_mat} consist of two modules. The first module ($i=1$) contains only nodes of degree $k=3$, and the second module ($i=2$) contains nodes of degrees $k=3$ and $k=11$ (see the schematic in Fig.~\ref{fig6_Network_sketch_Piikk}). From the first row of Eq.~\eqref{Piikk_mat}, we see that, on average, degree-3 nodes from module 1 have 20 connections to nodes of the same type (i.e., to degree-3 nodes in module 1) for every 1 edge to degree-11 nodes of module 2. Similarly, we see from row 3 that, on average, the degree-11 nodes (which only appear in module 2) have 1 edge to nodes of the same type and 1 edge to degree-3 nodes from module 1 for every 9 edges to degree-3 nodes from module 2. Finally, row 2 indicates that all degree-3 nodes from module 2 connect exclusively to degree-11 nodes.

The ensemble of $\Piikk$ networks described by the $\Piikk$ matrix~\eqref{Piikk_mat} cannot be accurately described by either the $\Pkk$ or $\Eii$ network models. We now demonstrate using two examples that $\Piikk$ theory is required to accurately analyze dynamics on such networks.

\begin{figure}[ht]
\centering
\includegraphics[width=0.97\columnwidth]{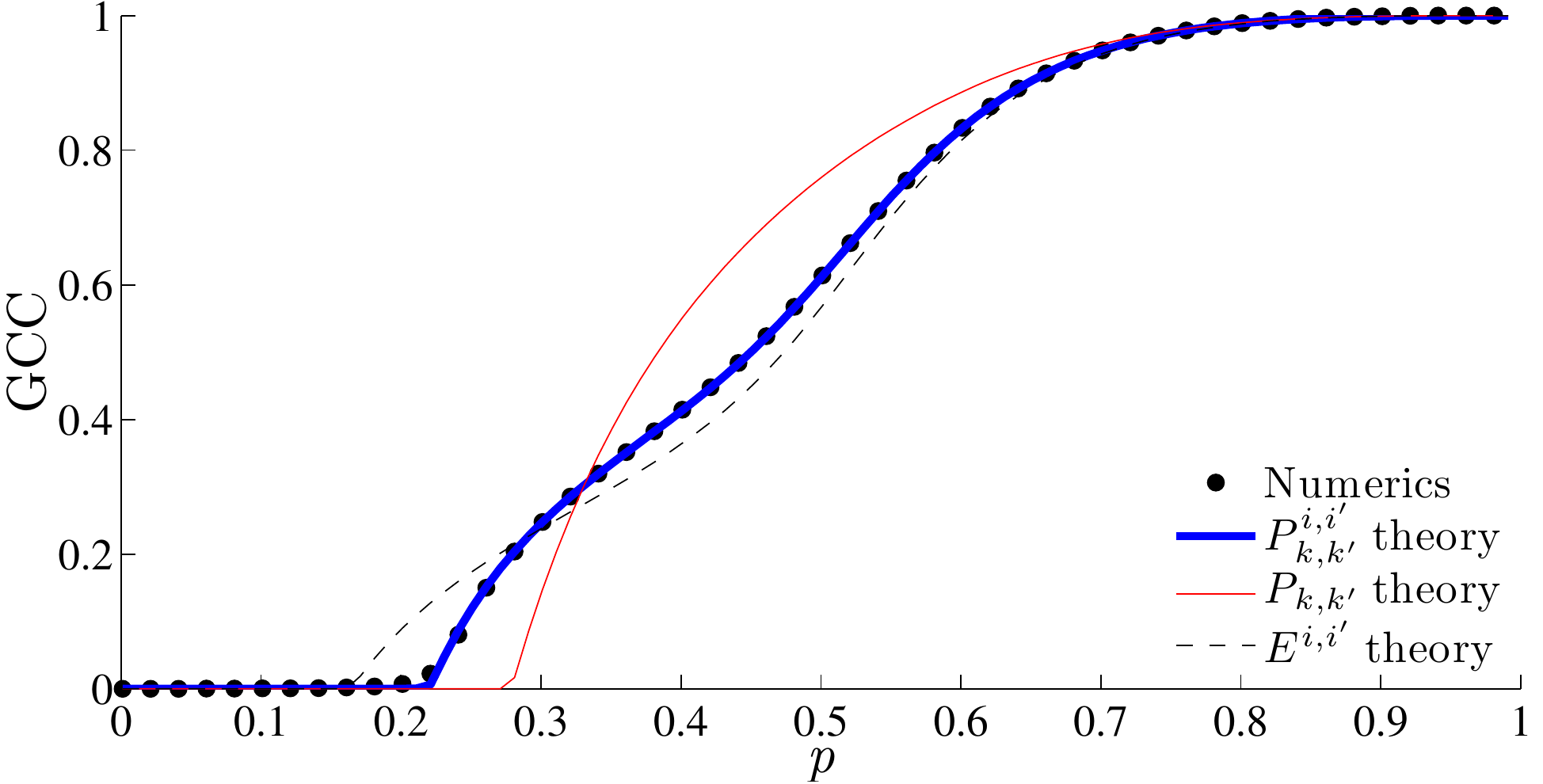}
\caption{GCC size versus bond occupation probability for a 25000-node $\Piikk$ network that consists of two modules generated according to the $\Piikk$ matrix~\eqref{Piikk_mat}. The first module consists only of nodes of degree 3, and the second module consists of nodes of degrees 3 and 11. The result from $\Piikk$ theory is indistinguishable from direct numerical simulations, whereas $\Pkk$ and $\Eii$ theories fail to accurately describe the observed behavior.}
\label{fig7_bperc_Piikk}
\end{figure}

In Fig.~\ref{fig7_bperc_Piikk}, we show bond-percolation results for a 25000-node $\Piikk$ network defined by the $\Piikk$ matrix in Eq.~\eqref{Piikk_mat} (cf.\ Fig.~\ref{fig3_bperc_Eii}). We calculate the $\Eii$ and $\Pkk$ distributions from~\eqref{Piikk_mat} using Eqs.~\eqref{relations}, and we plot in Fig.~\ref{fig7_bperc_Piikk} the GCC sizes that we calculate for each value of $p$ using $\Eii$, $\Pkk$, and $\Piikk$ theories. As one can see in the figure, only $\Piikk$ theory produces good agreement with the numerical simulations for all values of the bond-occupation probability. In particular, observe the differences in the predicted values of the percolation threshold $p_c$. Interestingly, $\Eii$ theory predicts a value of $p_c$ that is too low, whereas $\Pkk$ theory predicts a value that is too high.

\begin{figure}[t]
\centering
\includegraphics[width=0.97\columnwidth]{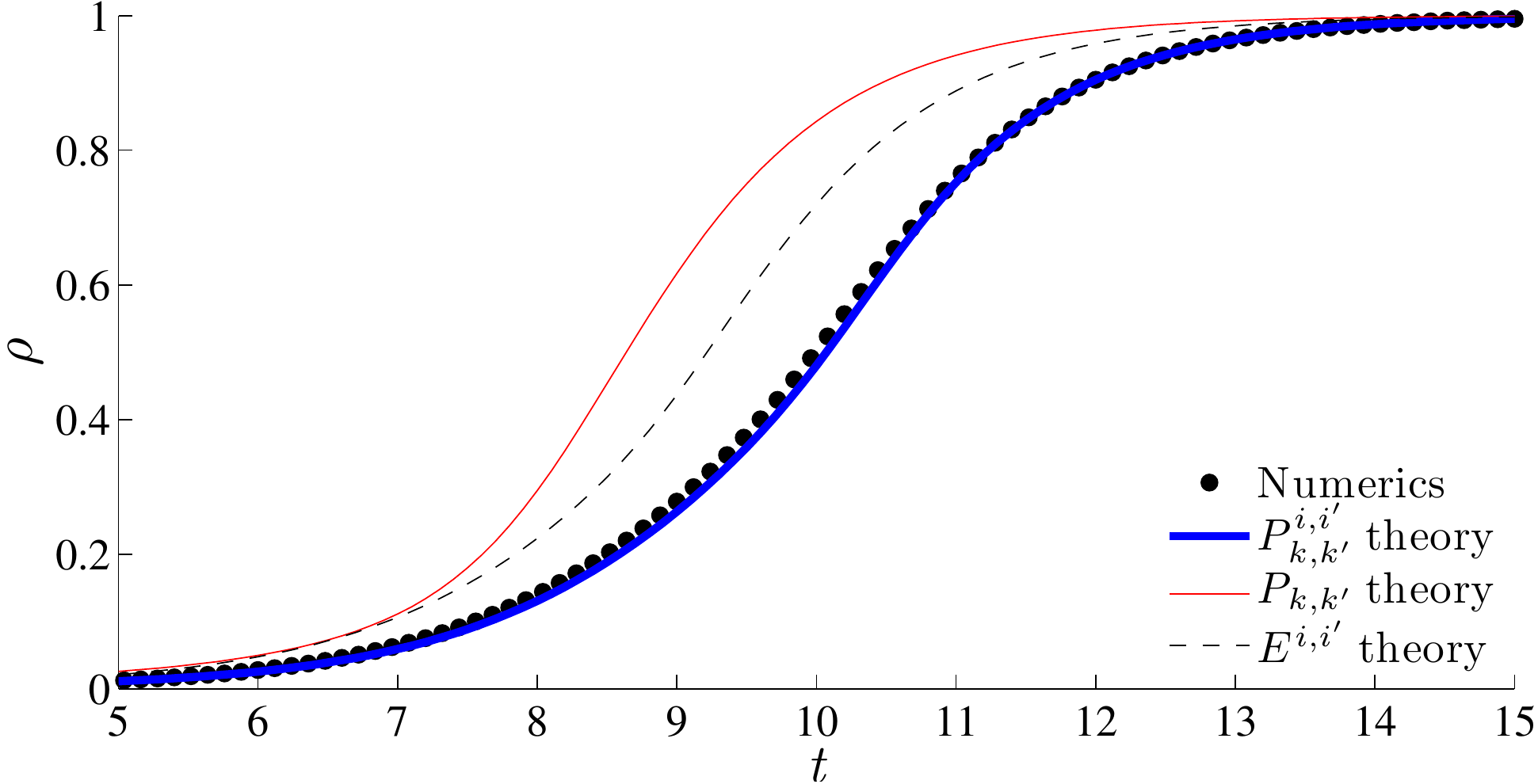}
\caption{Temporal evolution of fraction of active nodes in the Watts threshold model on a $\Piikk$ network with $N=5\times 10^5$ nodes distributed across two modules according to the $\Piikk$ matrix \eqref{Piikk_mat}. The first module consists only of nodes of degree 3, and the second module consists of nodes of degrees 3 and 11. The result of $\Piikk$ theory has better agreement with direct numerical simulation than do the results of $\Pkk$ and $\Eii$ theories.  Initially, we activate 0.1\% of nodes that are chosen uniformly at random from module 2 (or the equivalent fractions of nodes of each degree in the entire network for the case of $\Pkk$ theory). For our numerical computations, at each time step of size $\D t=1/N$, we update a single node chosen uniformly at random. The plotted results are averages over 10 realizations with different random seeds.}
\label{fig8_Watts_vs_t_Piikk}
\end{figure}

We use a network with the same $\Piikk$ matrix~\eqref{Piikk_mat} in Fig.~\ref{fig8_Watts_vs_t_Piikk} (but with a larger number of nodes) to demonstrate the temporal evolution of the fraction of active nodes in the Watts threshold model. Initially, we activate $0.1\%$ of nodes, which we chose uniformly at random from module 2. (To calculate the curve from $\Pkk$ theory, we set $\rho_k(0) = \sum_i \frac{N_i}{N} p_k^i \rho_k^i(0)$. This represents an initial activation of the same number of nodes of each degree as in the direct numerical simulations, but chosen uniformly at random from the entire network.) Nodes subsequently become active by interacting with their neighbors as specified by the response function~\eqref{F_Watts}. In this example, all nodes have the same threshold $R=0.18$. To obtain continuous-looking numerical results, we use asynchronous updating and update a single node (chosen uniformly at random) at each time step of size $\D t=1/N$. The theoretical curves also show dynamics in continuous time. [In the Appendix, we explain how to obtain them from the discrete-time Eqs.~\eqref{qikbar}--\eqref{rhoikn}]. Again, $\Piikk$ theory outperforms both $\Pkk$ and $\Eii$ theories.


\subsection{LFR Benchmark Networks}

We now consider a class of synthetic networks --- so-called ``LFR networks" --- that were developed by Lancichinetti, Fortunato, and Radicchi for benchmarking community-detection algorithms.~\cite{Lancichinetti08} LFR networks are an example of a planted-partition model,~\cite{condon2001} and they were developed to reflect important aspects of real-world networks better than previous such models.  Prominent features of LFR networks include power-law distributions of both node degrees and community sizes. In LFR networks, each node is assigned a degree from a power-law distribution with exponent $\gamma$ and cut-off degree $k_{\rm max}$.  Each node belongs to exactly one (planted) community, and a fraction $\mu$ of each node's edges are connected to nodes in other communities.  (The quantity $\mu$ is known as the ``mixing parameter''.) The number of nodes in each community is drawn from a power-law distribution with exponent $\beta$, such that the total number of nodes in the network is $N$. The minimum and maximum permitted community sizes are specified, respectively, by the parameters $c_{\rm min}$ and $c_{\rm max}$. 

Because the connections between nodes in LFR networks are independent of node degrees, the $\Eii$ network model (which is a special case of our $\Piikk$ network model) should be sufficient to describe these networks. However, due to additional constraints imposed on LFR networks~\cite{Lancichinetti08} they are atypical examples of the $\Eii$ ensemble. Specifically, in LFR networks the distribution of the number of external edges of a degree-$k$ node (i.e., the number of its neighbors that belong to other communities) is highly peaked near $\mu k$ (indeed, a delta function if $\mu k$ is an integer), whereas the corresponding $\Eii$ ensemble instead has a binomial distribution with mean $\mu k$.

\begin{figure}[htb]
\centering
\includegraphics[width=0.97\columnwidth]{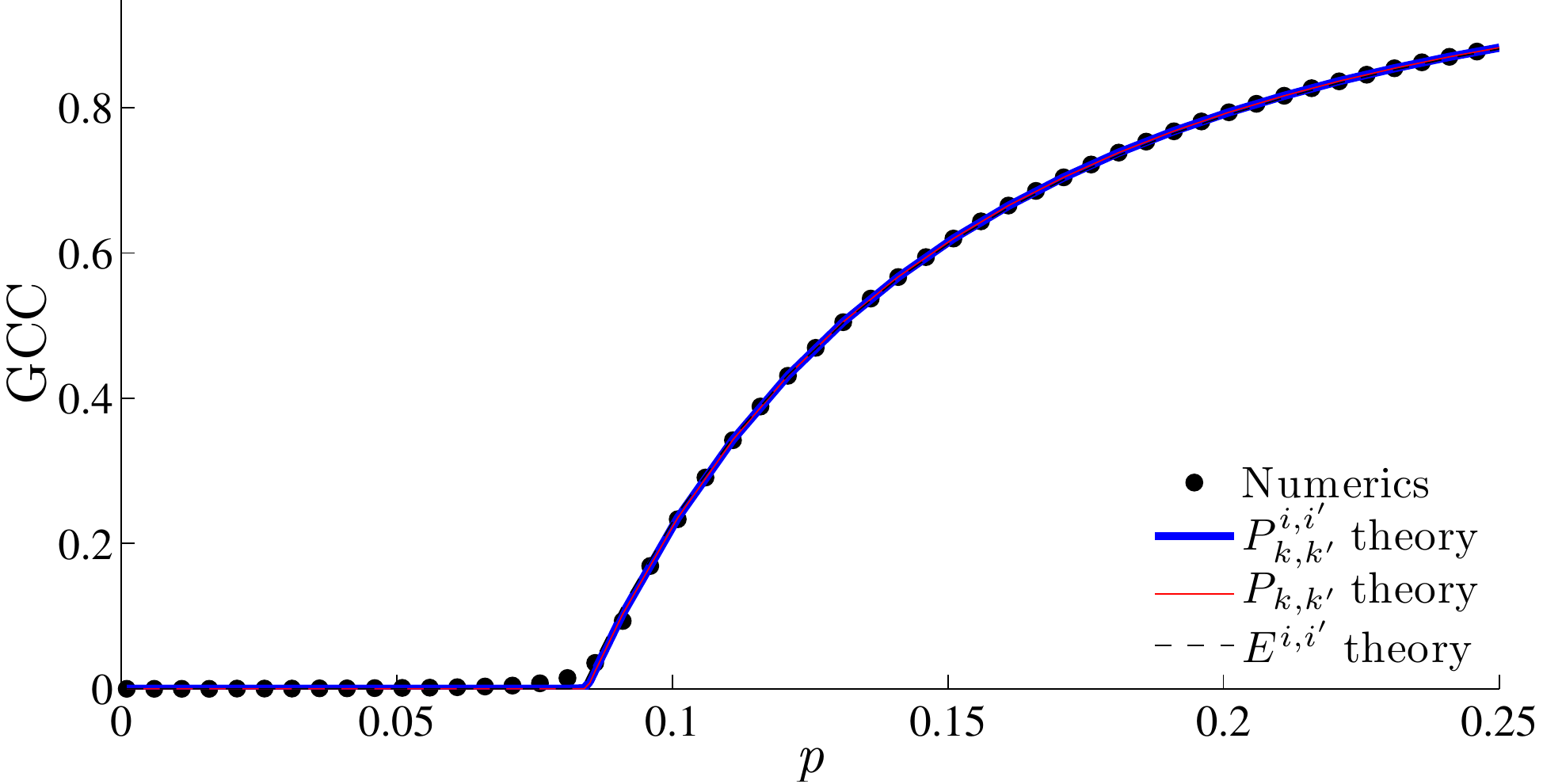}
\caption{GCC size versus bond-occupation probability for an LFR network constructed using the following parameters: $N=25000$, $z=10$, $k_{\rm max}=30$, $c_{\rm min}=2500$, $c_{\rm max}=5000$, $\gamma=-2.5$, $\beta = -1.5$, and $\mu = 0.2$. All theoretical curves predict the numerical behavior extremely well.}
\label{fig_bperc_LFR}
\end{figure}

Furthermore, because nodes in different communities in LFR networks have degrees that are drawn from the same power-law distribution, the single-module $\Pkk$ theory becomes adequate for LFR networks for some processes for which the modular structure is not very important. We demonstrate this phenomenon using the example of bond percolation (see Fig.~\ref{fig_bperc_LFR}). In our examples, we extract the $\Pkk$, $\Eii$, and $\Piikk$ distributions from the adjacency matrix of each generated LFR network and use them in Eqs.~\eqref{qikbar}--\eqref{rhoikn} to predict the results of dynamical processes on these networks.

In Fig.~\ref{fig_bperc_LFR}, we show bond-percolation results on an LFR network that is constructed using the parameters specified in the caption. All theories accurately predict the GCC size as we vary the bond-occupation probability $p$. This result is expected, because the modules in an LFR network are essentially scaled copies of each other (i.e., they are statistically similar to each other), as they have the same degree distribution and inter-module connections that are drawn from the same random process. Therefore, as long as there are sufficiently many inter-module edges (to help avoid finite-size effects), even $\Pkk$ theory can correctly predict the GCC size.

\begin{figure}[htb]
\centering
\includegraphics[width=0.97\columnwidth]{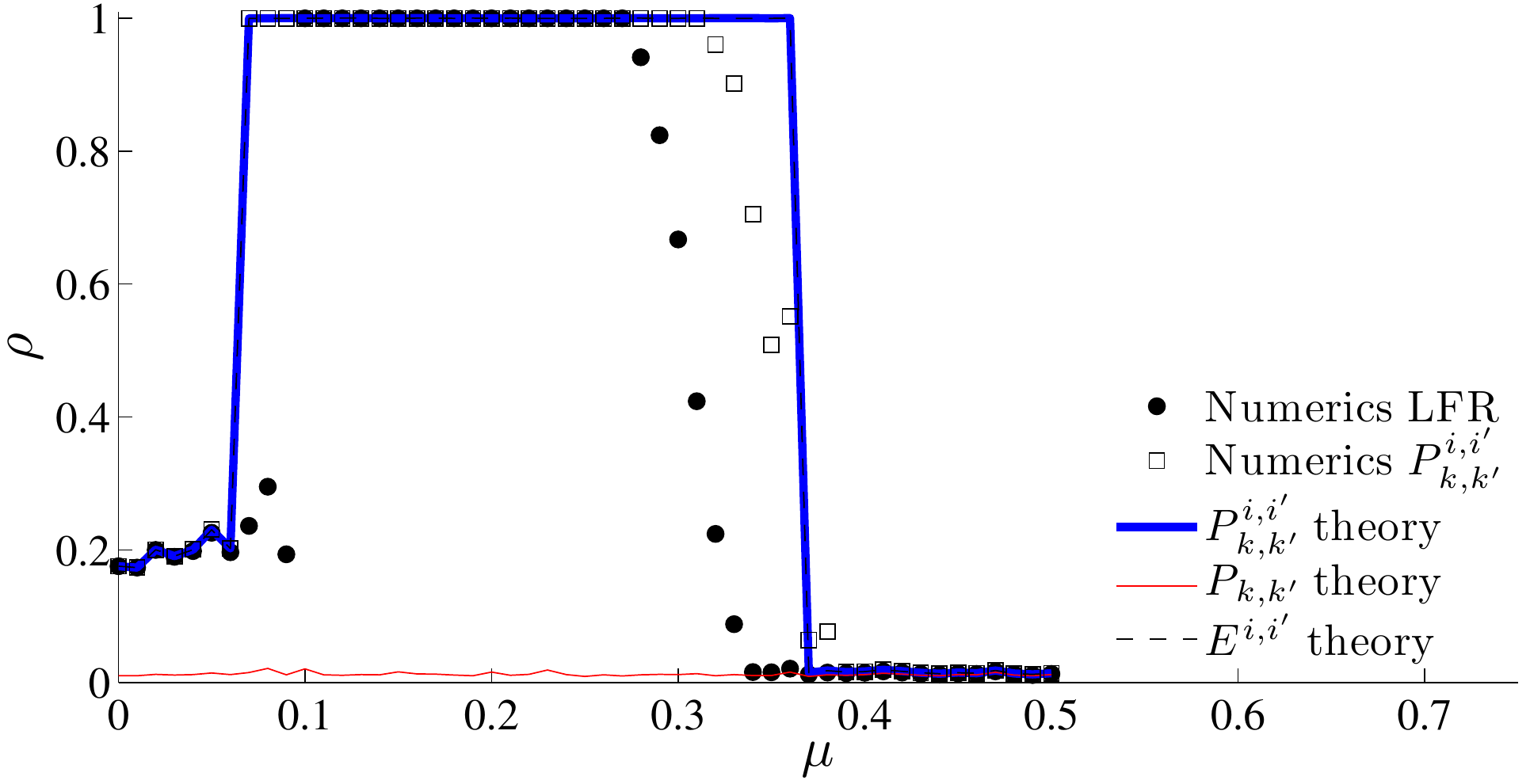}
\caption{Final fraction of active nodes for the Watts threshold model on LFR networks as a function of the mixing parameter $\mu$. The other parameters are the same as in Fig.~\ref{fig_bperc_LFR}. All nodes have the same threshold $R=0.18$, and 5\% of the nodes (chosen uniformly at random) in the largest community are initially active. 
For our numerical computations, for each value of $\mu$ we average over 100 choices of random seeds on a single realization of an LFR (or $\Piikk$) network.}
\label{fig_Watts_vs_mu_LFR}
\end{figure}

We now consider an example of the Watts threshold model on LFR networks in which we initially activate 5\% of nodes in the largest module. In Fig.~\ref{fig_Watts_vs_mu_LFR}, we show the final fraction of active nodes versus the mixing parameter $\mu$.  We observe two transitions: one occurs for small $\mu$, and the other occurs for large $\mu$. For small values of $\mu$, the final activation size is limited by the size of the module that contains the seed nodes. Hence, the small-$\mu$ transition to a cascade occurs when a network becomes sufficiently connected so that the activation from the seed module can propagate to other modules. The large-$\mu$ transition away from the cascade occurs when the nodes in the seed module have enough edges to inactive nodes in other modules so that they can no longer activate. The activation of nodes thereby does not propagate even within the module that contains the seed nodes. A similar phenomenon was recently reported in Ref.~\onlinecite{Nematzadeh14} using a different ensemble of random networks.

The results of $\Piikk$ and $\Eii$ theories are indistinguishable from each other, and both do a reasonable (but clearly imperfect) job at describing the observed numerical behavior on LFR networks. We attribute the differences between our theory and the simulations on LFR networks to the finite network size --- recall that our theory assumes an infinitely large network --- and to the differences between the LFR and $\Eii$ network ensembles that we highlighted above. To estimate the influence of the second effect, we perform numerical simulations on $\Piikk$ networks that we generate using the $\Piikk$ matrix of the corresponding LFR networks. In other words, we generate an LFR network, calculate its $\Piikk$ matrix, and then use this matrix to create the corresponding $\Piikk$ network.~\cite{Piikk_url} The numerical simulation results on such $\Piikk$ networks are described by the theory very well. The curve from $\Pkk$ theory predicts virtually no propagation of activation beyond seed nodes for all values of $\mu$ because it ignores modular structure in networks. Therefore, $\Pkk$ theory only gives a good prediction for the values of $\mu$ that are above the large-$\mu$ (i.e., downward) transition in Fig.~\ref{fig_Watts_vs_mu_LFR}.


\subsection{Conjoined Real-World Networks}

Thus far, we have considered synthetic networks that are generated using the $\Piikk$ model (and special cases thereof). It is also important to examine the applicability of $\Piikk$ theory to real-world networks. As a first example, we take two real-world networks and conjoin them to form a single network by adding random edges between them such that the original networks can be considered as modules in the resulting network.

\begin{figure}[htb]
\centering
\includegraphics[width=0.97\columnwidth]{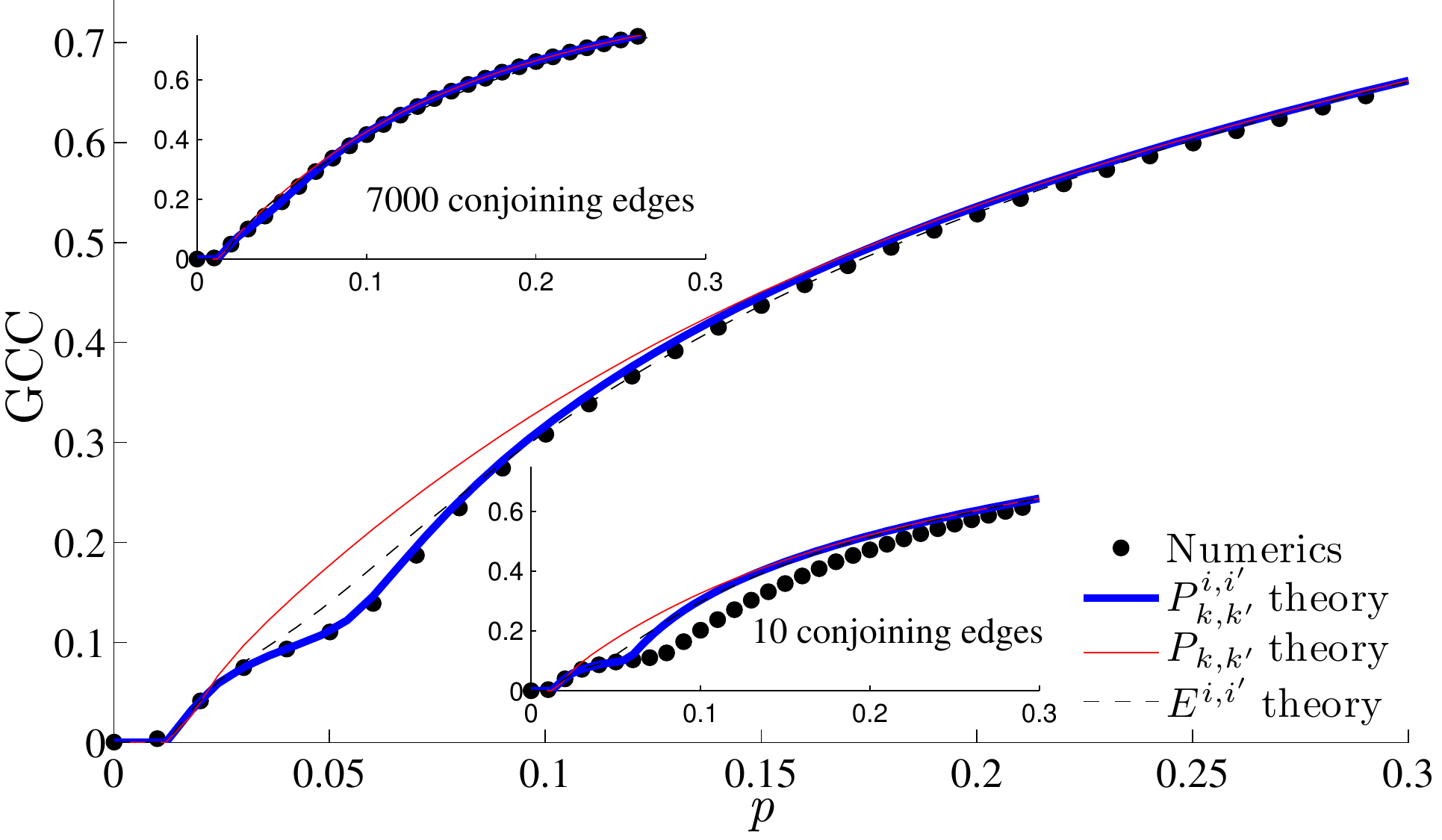}
\caption{GCC size versus bond-occupation probability for the Caltech Facebook network conjoined to a protein interaction network. The results of $\Piikk$ theory are better than those for $\Pkk$ and $\Eii$ theories. In this example, we added 700 edges uniformly at random between the first network and the second. The insets show the results for using many more (7000) or far fewer (10) conjoining edges. When there are 7000 conjoining edges, the resulting network becomes essentially random and all three theories become adequate. When there are too few conjoining edges, finite-size effects prevent any of the three theories from working well.}
\label{fig9_bperc_Caltech_DIP}
\end{figure}

In Fig.~\ref{fig9_bperc_Caltech_DIP}, we examine the bond-percolation properties of a network that we construct by conjoining a Facebook network of Caltech~\cite{Traud08,Traud12} to a network of interacting proteins~\cite{Colizza05,Colizza06,DIP_url_both} using 700 edges (which we call ``conjoining edges") that we add uniformly at random between the two networks, subject to the restriction that the added edges lead exclusively from one network to the other.~\footnote{Admittedly, this is not the most realistic example of interacting networks, but we use it because it illustrates our conceptual point. One can make similar comparisons using more complicated algorithms to conjoin the Caltech and protein-interaction networks. For example, the conjoining edges could depend on node degrees, modular structure, or other properties.}
We then apply $\Pkk$, $\Eii$, and $\Piikk$ theories to see how well they capture the results of direct numerical simulations of bond percolation on the conjoined networks.~\footnote{For $\Piikk$ or $\Eii$ theory to work, we need to conjoin networks using sufficiently many edges to avoid finite-size effects, which are not captured by the theories.}
Unlike the synthetic examples that we discussed previously, the conjoined Caltech-protein network has naturally-occurring degree-degree correlations inside of each module. The results of Fig.~\ref{fig9_bperc_Caltech_DIP} demonstrate that $\Piikk$ theory captures the bond-percolation properties better than the other theories.

\begin{figure}[htb]
\centering
\includegraphics[width=0.97\columnwidth]{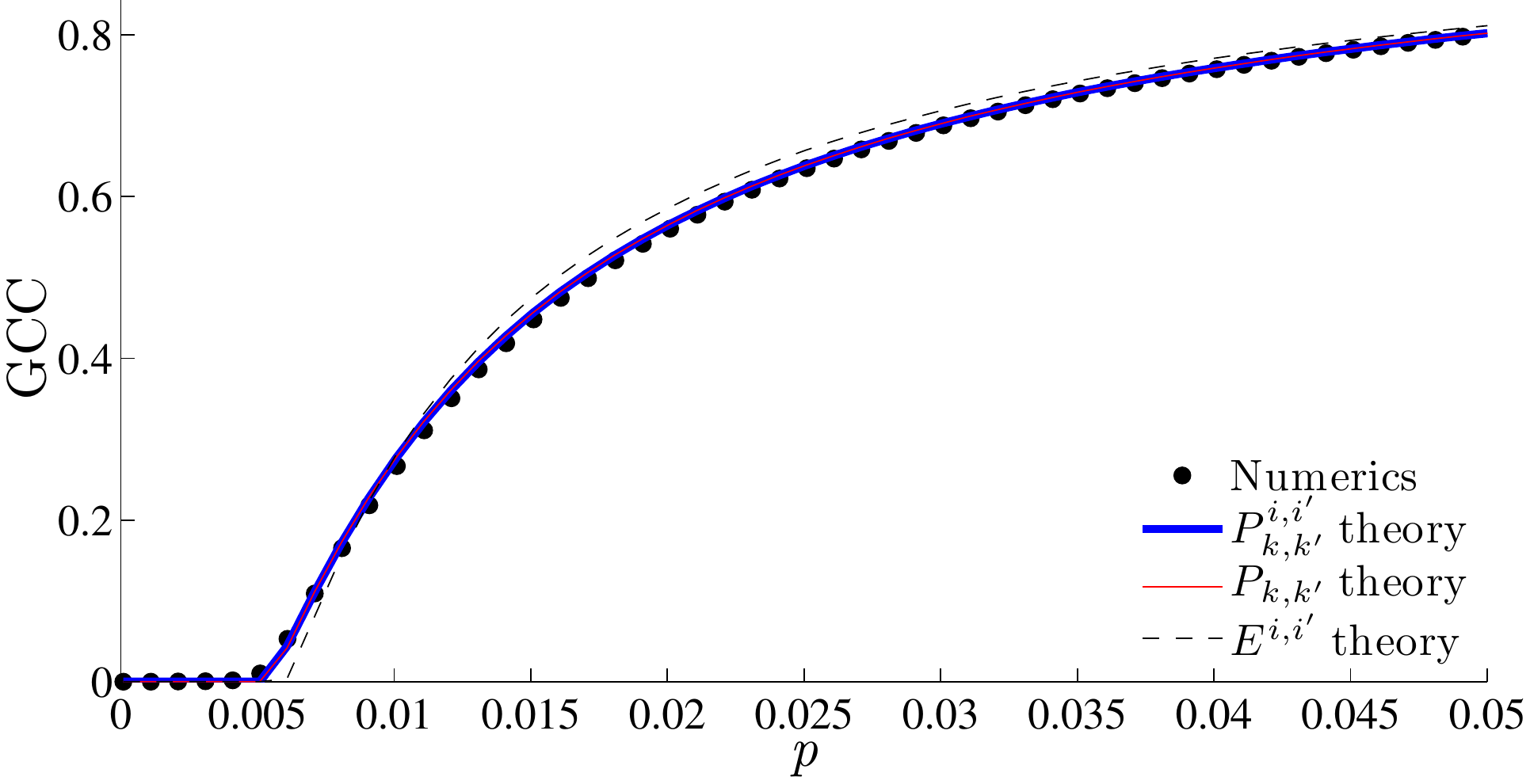}
\caption{GCC size versus bond-occupation probability for a multi-university Facebook network. All theories predict the GCC size very accurately. The results of $\Piikk$ and $\Pkk$ theories are slightly better than those of $\Eii$ theory. This suggests that degree-degree correlations play a more important role than modular structure in this example.}
\label{fig_bperc_FB_MI}
\end{figure}

In Fig.~\ref{fig_bperc_FB_MI}, we show bond-percolation results for a multi-university Facebook network. This provides a real-world example of interdependent networks rather than two disparate real-world networks that we conjoined artificially. We consider the Michigan23 and MSU24 Facebook networks~\cite{Traud12} along with the inter-university edges that exist between these users (i.e., nodes) from the different universities. The largest connected component (LCC) of the aggregate network has $N=62770$ nodes, a mean degree of $z\approx82$, a degree-degree Pearson-correlation coefficient (i.e., degree assortativity) of $r\approx0.044$, and a clustering coefficient~\cite{Watts98} of $C\approx 0.201$. The number of inter-university edges is $m=290278$ (i.e., about $11.2\%$ of the total number of edges). When separated, the LCCs of the Michigan23 and MSU24 modules of the network have, respectively, $N=30106$ and $N=32361$ nodes, mean degrees of $z\approx78$ and $z\approx69$, degree-degree Pearson-correlation coefficients of $r\approx0.115$ and $r\approx0.009$, and clustering coefficients of $C\approx0.21$ and $C\approx0.204$.

In Fig.~\ref{fig_bperc_FB_MI}, we compare the predictions from $\Piikk$, $\Eii$, and $\Pkk$ theories. As one can see in the figure, all theories adequately predict the GCC size. Interestingly, $\Eii$ theory performs slightly worse than the other theories. This suggests that modules in this network have nontrivial degree-degree correlations that seem to play a more important role than modular structure in this example.

\begin{figure}[ht]
\centering
\includegraphics[width=0.97\columnwidth]{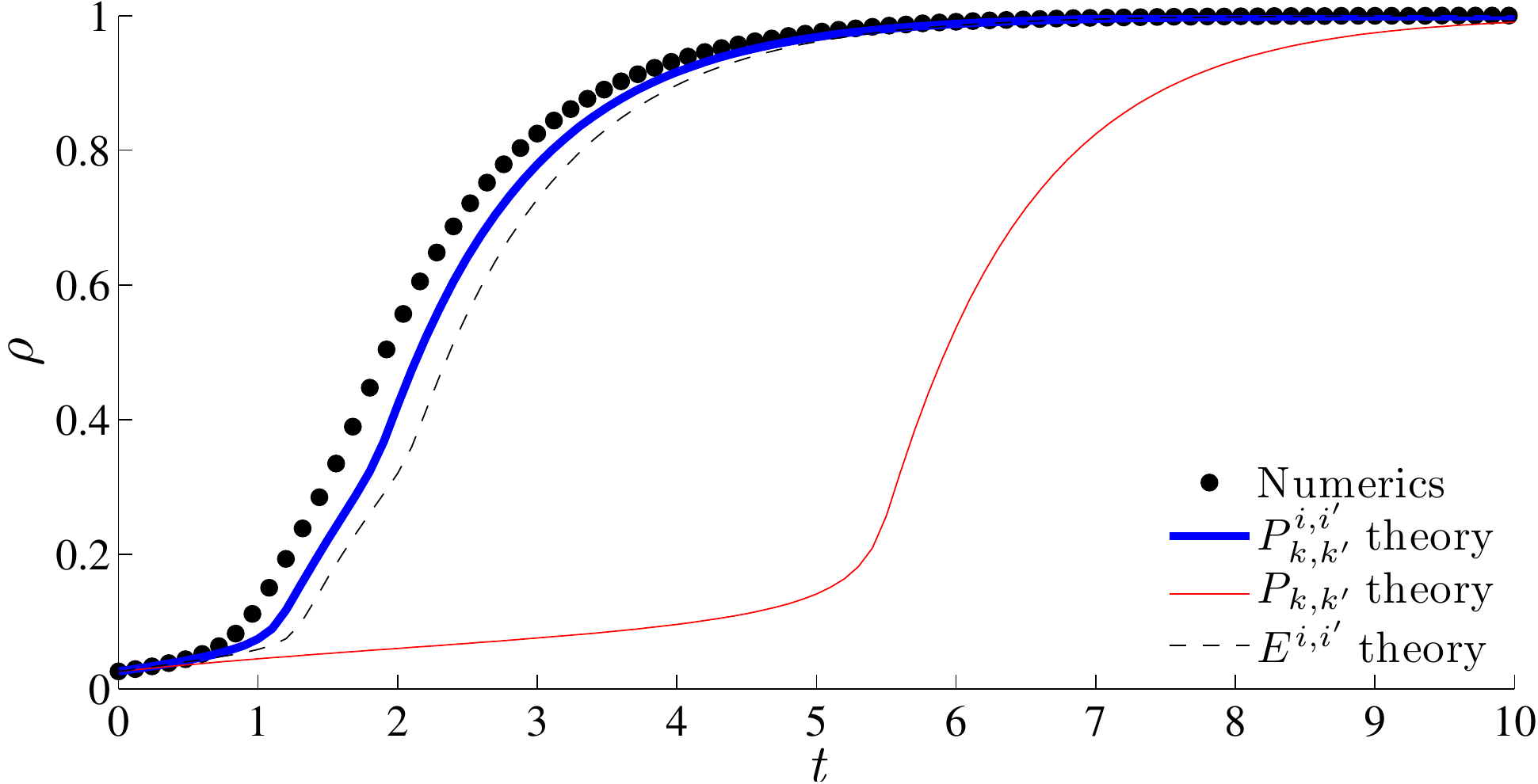}\put(-210,110){\text{(a)}}\\
\includegraphics[width=0.97\columnwidth]{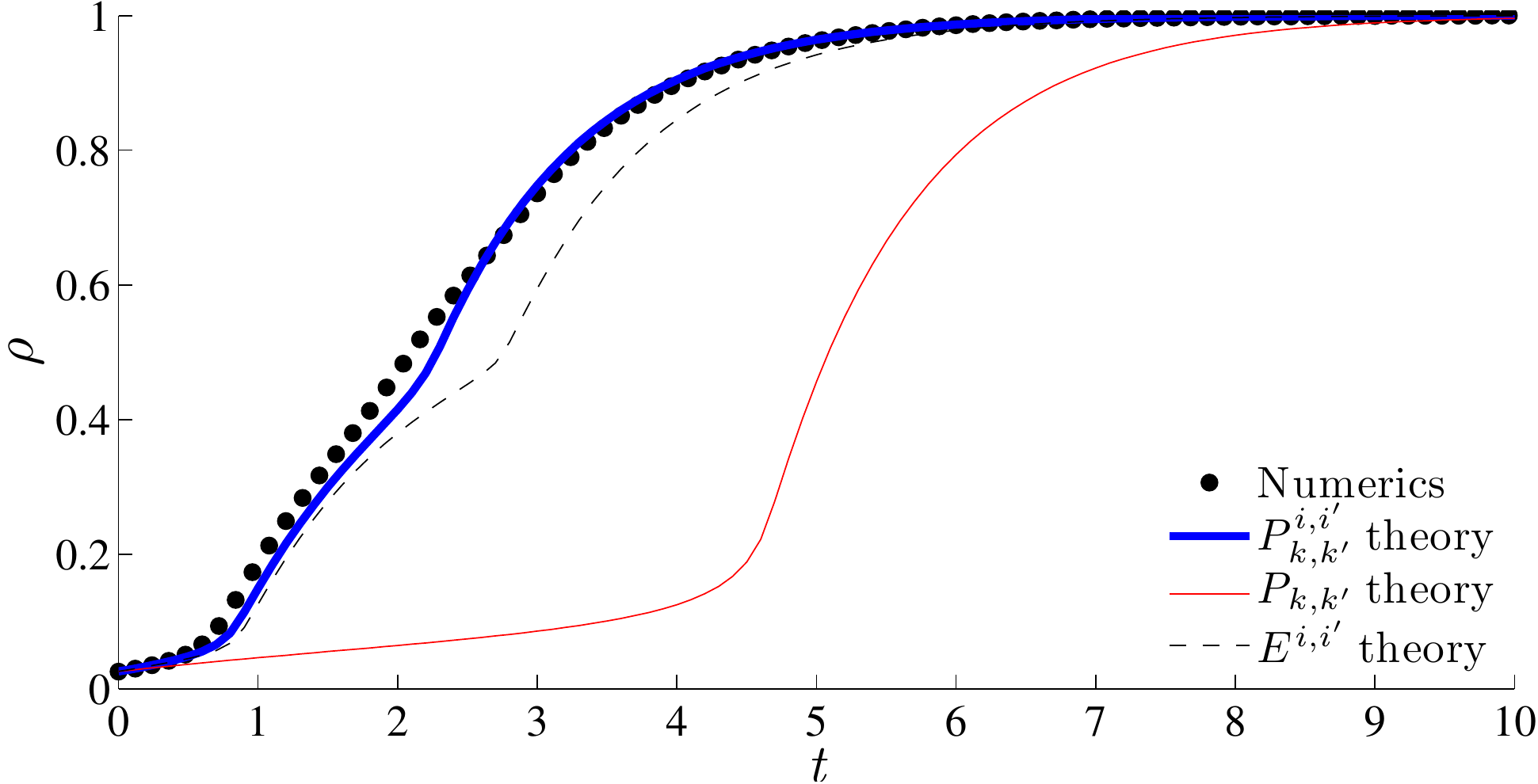}\put(-210,110){\text{(b)}}\\
\includegraphics[width=0.97\columnwidth]{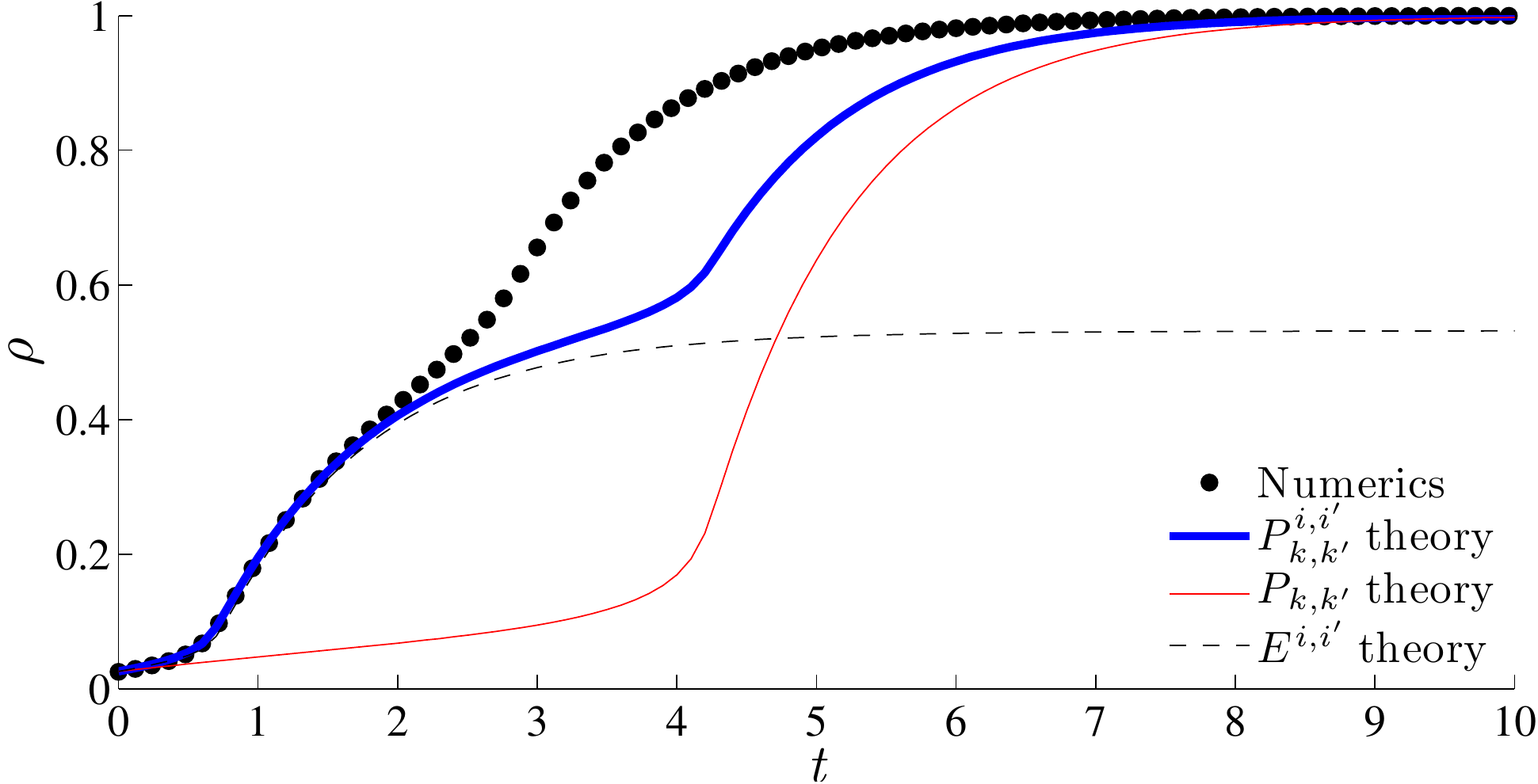}\put(-210,110){\text{(c)}}
\caption{(a) Temporal evolution of the fraction of active nodes in the Watts threshold model on a multi-university Facebook network. We update nodes asynchronously by choosing a single node uniformly at random during each time step of size $\D t=1/N$. All nodes have an identical threshold value of $R=0.1$, and we initially activate (uniformly at random) $5\%$ of the nodes in the MSU24 module. For $\Pkk$ theory, we use a seed fraction that gives the same number of nodes of each degree, but chosen (uniformly at random) from the entire network. We average the results of our numerical computations over 10 choices of random seeds. Panels (b) and (c) are the same as in panel (a), but we additionally remove (b) 50\% and (c) 80\% of the inter-university edges uniformly at random.}
\label{fig10_Watts_vs_t_FB_MI}
\end{figure}

In Fig.~\ref{fig10_Watts_vs_t_FB_MI}, we show continuous-time evolution of the fraction of active nodes for the Watts threshold model on the same multi-university Facebook network that we used in Fig.~\ref{fig_bperc_FB_MI}. As an initial condition, we select (uniformly at random) 5\% of the nodes in the MSU24 module to be initially active. (For $\Pkk$ theory, we set $\rho_k(0) = \sum_i \frac{N_i}{N} p_k^i \rho_k^i(0)$, which represents the initial activation of the same number of nodes of each degree as in the simulated case, but we select the nodes uniformly at random from the entire network.) At each subsequent time step (of size $\D t = 1/N$), we update a single node (chosen uniformly at random) according to the threshold rules~\eqref{F_Watts}. As in previous examples, one can see in Fig.~\ref{fig10_Watts_vs_t_FB_MI} that $\Piikk$ theory outperforms $\Pkk$ and $\Eii$ theories. For example, in Fig.~\ref{fig10_Watts_vs_t_FB_MI}(c), only $\Piikk$ theory accurately predicts the initial temporal evolution and the final fraction of active nodes. The other two theories either depart significantly away from the numerical curve or incorrectly predict the final fraction of active nodes.

Although $\Piikk$ theory does better than both $\Pkk$ theory and $\Eii$ theory on this example, none of the three theoretical curves fit the numerical results particularly well. Every theory is of course limited, and it has also previously been demonstrated that the Watts threshold model with identical thresholds on real-world networks is less amenable to analytical treatments of the style that we use in this paper than, e.g., site and bond percolation.  See Refs.~\onlinecite{Melnik11,Gleeson12} for details. We purposely show this example because we know it is very challenging for these theories, and it is therefore insightful to see how a more intricate theory performs on it. Interestingly, all theories predict slower activation than what we observe numerically in Fig.~\ref{fig10_Watts_vs_t_FB_MI}. In contrast, $\Pkk$ and $\Eii$ theories predict faster activation in the example that we illustrated in Fig.~\ref{fig8_Watts_vs_t_Piikk}.  There are many possible reasons that this can occur --- including the differences in degree distribution, degree-degree correlations, density of loops of different lengths, community structure, and so on.


\section{Conclusions} \label{sec7}

We have developed a new random-graph ensemble to describe multi-module networks in which modules can have different degree-degree correlations and the connections between modules are also defined by the joint degree-degree distribution of nodes for each pair of modules. We also presented an analytical method for analyzing a broad class of binary-state dynamics on such networks. Our $\Piikk$ network model generalizes the configuration model, random networks with degree-degree correlations (i.e., $\Pkk$ networks), and multi-module networks that have no a priori degree-degree correlations (e.g., $\Eii$ networks and the LFR benchmark networks), and it provides an alternative description of interacting networks to those that have been examined recently by other authors.~\cite{Kivela13,Leicht09,Allard09,Brummitt10,Mendiola12}

We have also demonstrated using both synthetic networks and real-world networks that $\Piikk$ theory can explain dynamics that neither $\Pkk$ nor $\Eii$ models are able to capture. In particular, the analytical approach that we presented allows one to consider situations in which dynamics can be different on different modules. We provide code online~\cite{Piikk_url} for generating networks from our $\Piikk$ ensemble. 
For example, this should be useful for benchmarking community-detection methods. Our model can also be generalized to include more than two (degree and module) types of correlations, and we expect that such efforts will be important for the investigation of dynamics on complicated interdependent networks.


\section*{Acknowledgements}

S.M. acknowledges the INSPIRE fellowship funded by the Irish Research Council (co-funded by Marie Curie Actions under FP7). J.P.G. and S.M. acknowledge funding provided by Science Foundation Ireland under programmes 11/PI/1026. M.A.P. acknowledges a research award (\#220020177) from the James S. McDonnell Foundation. M.A.P and J.P.G. were supported by the European Commission FET-Proactive project PLEXMATH (Grant No.317614). P.J.M. was funded by Award Number R21GM099493 from the National Institute of General Medical Sciences; the content is solely the responsibility of the authors and does not necessarily represent the official views of the National Institute of General Medical Sciences or the National Institutes of Health. We thank Ali Faqeeh for useful discussions. We thank Adam D'Angelo and Facebook for providing the Facebook data. We also thank Cx-Nets collaboratory for making publicly available the protein interaction data set that we used in this study. This work was conceived in part during the 2010--11 program year on Complex Networks at the Statistical and Applied Mathematical Sciences Institute (SAMSI) in Research Triangle Park, NC, USA.


\appendix*
\section{Continuous-Time Analytical Approximation}

In this appendix, we demonstrate how the discrete-time Eqs.~\eqref{qikbar}--\eqref{rhoikn} can be used to approximate continuous-time evolution.~\cite{Gleeson08a} We first rewrite Eqs.~\eqref{qikn} and~\eqref{rhoikn} as
\begin{align}
\label{qg}
    q^i_k(n+1) &\equiv g_k^i\lb( \bar q_k^i(n) \rb)\,,\\
\nn \rho_k^i(n+1) &\equiv h_k^i\lb( \bar q_k^i(n) \rb)\,,
\end{align}
where $\bar q^{i}_k(n)$ is obtained from Eq.~\eqref{qikbar}. These equations describe the case of synchronous updating, in which the states of all $N$ network nodes are updated at each discrete time step $n$. It is possible to modify these equations to account for situations in which only the states of a certain fraction $\tau$ of nodes (chosen uniformly at random) are updated. Thus, the value $\tau=1$ corresponds to synchronous updating of all nodes, and $\tau=1/N$ corresponds to the completely asynchronous case in which a single node (chosen uniformly at random) is updated at each time step. For the monotonic dynamical processes that we consider in this paper, both types of updating lead to the same final state. However, the transient dynamics can be different.

To deal with asynchronous updating, in which only a fraction $\tau$ of nodes is updated at each time step, we use a time step of $\D t=\tau$ so that we have a common time scale for all $\tau$ (including the synchronous updating case of $\tau=1$). If the updating is synchronous (i.e., if $\tau=1$) as in Eqs.~\eqref{qg}, then the probability $q^i_k$ increases by $\D q^i_k = g_k^i\lb(\bar q_k^i \rb) - q^i_k$. In other words, all nodes that are available for activation are activated. In the asynchronous updating case, only a fraction $\tau$ of all nodes available for activation are activated, so $q^i_k$ increases by $\D q^i_k = \tau \lb(g_k^i\lb( \bar q_k^i \rb) - q^i_k\rb)$. Therefore, for sufficiently low values of $\tau$, it is permissible to approximate the temporal evolutions of $q^i_k$ and $\rho_k^i$ as continuous. This yields the following set of ordinary differential equations:
\begin{align}
\label{cont_time}
    \frac{d q^i_k(t)}{dt} &= g_k^i\lb( \bar q_k^i(t) \rb) - q^i_k(t)\,,\\
\nn \frac{d\rho_k^i(t)}{dt} &= h_k^i\lb( \bar q_k^i(t) \rb) - \rho_k^i(t)\,,
\end{align}
We use Eqs.~\eqref{cont_time} to produce continuous-theory curves for the Watts threshold model in Figs.~\ref{fig8_Watts_vs_t_Piikk} and~\ref{fig10_Watts_vs_t_FB_MI} in the main text.


\bibliographystyle{aipnum4-1}
\bibliography{networks}

\begin{thebibliography}{10}%
\makeatletter
\providecommand \@ifxundefined [1]{%
 \ifx #1\undefined \expandafter \@firstoftwo
 \else \expandafter \@secondoftwo
\fi
}%
\providecommand \@ifnum [1]{%
 \ifnum #1\expandafter \@firstoftwo
 \else \expandafter \@secondoftwo
\fi
}%
\providecommand \enquote [1]{``#1''}%
\providecommand \bibnamefont  [1]{#1}%
\providecommand \bibfnamefont [1]{#1}%
\providecommand \citenamefont [1]{#1}%
\providecommand\href[0]{\@sanitize\@href}%
\providecommand\@href[1]{\endgroup\@@startlink{#1}\endgroup\@@href}%
\providecommand\@@href[1]{#1\@@endlink}%
\providecommand \@sanitize [0]{\begingroup\catcode`\&12\catcode`\#12\relax}%
\@ifxundefined \pdfoutput {\@firstoftwo}{%
 \@ifnum{\z@=\pdfoutput}{\@firstoftwo}{\@secondoftwo}%
}{%
 \providecommand\@@startlink[1]{\leavevmode\special{html:<a href="#1">}}%
 \providecommand\@@endlink[0]{\special{html:</a>}}%
}{%
 \providecommand\@@startlink[1]{%
  \leavevmode
  \pdfstartlink
   attr{/Border[0 0 1 ]/H/I/C[0 1 1]}%
   user{/Subtype/Link/A<</Type/Action/S/URI/URI(#1)>>}%
  \relax
 }%
 \providecommand\@@endlink[0]{\pdfendlink}%
}%
\providecommand \url  [0]{\begingroup\@sanitize \@url }%
\providecommand \@url [1]{\endgroup\@href {#1}{\urlprefix}}%
\providecommand \urlprefix [0]{URL }%
\providecommand \Eprint[0]{\href }%
\@ifxundefined \urlstyle {%
  \providecommand \doi [1]{doi:\discretionary{}{}{}#1}%
}{%
  \providecommand \doi [0]{doi:\discretionary{}{}{}\begingroup
  \urlstyle{rm}\Url }%
}%
\providecommand \doibase [0]{http://dx.doi.org/}%
\providecommand \Doi[1]{\href{\doibase#1}}%
\providecommand \selectlanguage [0]{\@gobble}%
\providecommand \bibinfo [0]{\@secondoftwo}%
\providecommand \bibfield [0]{\@secondoftwo}%
\providecommand \translation [1]{[#1]}%
\providecommand \BibitemOpen[0]{}%
\providecommand \bibitemStop [0]{}%
\providecommand \bibitemNoStop [0]{.\EOS\space}%
\providecommand \EOS [0]{\spacefactor3000\relax}%
\providecommand \BibitemShut [1]{\csname bibitem#1\endcsname}%
\bibitem{comnotices}%
  \BibitemOpen
  \bibfield{author}{%
  \bibinfo {author} {\bibfnamefont{M.~A.}\ \bibnamefont{Porter}}, \bibinfo
  {author} {\bibfnamefont{J.-P.}\ \bibnamefont{Onnela}},\ and\ \bibinfo
  {author} {\bibfnamefont{P.~J.}\ \bibnamefont{Mucha}},\ }%
  \bibfield{journal}{%
  \bibinfo {journal} {Not. Amer. Math. Soc.}\ }%
  \textbf{\bibinfo {volume} {56}},\ \bibinfo {pages} {1082} (\bibinfo {year}
  {2009})\BibitemShut{NoStop}%
\bibitem{Fortunato10}%
  \BibitemOpen
  \bibfield{author}{%
  \bibinfo {author} {\bibfnamefont{S.}~\bibnamefont{Fortunato}},\ }%
  \bibfield{journal}{%
  \bibinfo {journal} {Phys. Rep.}\ }%
  \textbf{\bibinfo {volume} {486}},\ \bibinfo {pages} {75} (\bibinfo {year}
  {2010})\BibitemShut{NoStop}%
\bibitem{Kivela13}%
  \BibitemOpen
  \bibfield{author}{%
  \bibinfo {author} {\bibfnamefont{M.}~\bibnamefont{Kivel\"{a}}}, \bibinfo
  {author} {\bibfnamefont{A.}~\bibnamefont{Arenas}}, \bibinfo {author}
  {\bibfnamefont{M.}~\bibnamefont{Barthelemy}}, \bibinfo {author}
  {\bibfnamefont{J.~P.}\ \bibnamefont{Gleeson}}, \bibinfo {author}
  {\bibfnamefont{Y.}~\bibnamefont{Moreno}},\ and\ \bibinfo {author}
  {\bibfnamefont{M.~A.}\ \bibnamefont{Porter}},\ }%
  \bibfield{journal}{%
  \bibinfo {journal} {arXiv}\ }%
  \textbf{\bibinfo {volume} {1309.7233}} (\bibinfo {year}
  {2013})\BibitemShut{NoStop}%
\bibitem{Vespignani10}%
  \BibitemOpen
  \bibfield{author}{%
  \bibinfo {author} {\bibfnamefont{A.}~\bibnamefont{Vespignani}},\ }%
  \bibfield{journal}{%
  \bibinfo {journal} {Nature (London)}\ }%
  \textbf{\bibinfo {volume} {464}},\ \bibinfo {pages} {984} (\bibinfo {year}
  {2010})\BibitemShut{NoStop}%
\bibitem{Mucha10}%
  \BibitemOpen
  \bibfield{author}{%
  \bibinfo {author} {\bibfnamefont{P.~J.}\ \bibnamefont{Mucha}}, \bibinfo
  {author} {\bibfnamefont{T.}~\bibnamefont{Richardson}}, \bibinfo {author}
  {\bibfnamefont{K.}~\bibnamefont{Macon}}, \bibinfo {author}
  {\bibfnamefont{M.~A.}\ \bibnamefont{Porter}},\ and\ \bibinfo {author}
  {\bibfnamefont{J.-P.}\ \bibnamefont{Onnela}},\ }%
  \bibfield{journal}{%
  \bibinfo {journal} {Science}\ }%
  \textbf{\bibinfo {volume} {328}},\ \bibinfo {pages} {876} (\bibinfo {year}
  {2010})\BibitemShut{NoStop}%
\bibitem{Zhou06}%
  \BibitemOpen
  \bibfield{author}{%
  \bibinfo {author} {\bibfnamefont{C.}~\bibnamefont{Zhou}}, \bibinfo {author}
  {\bibfnamefont{L.}~\bibnamefont{Zemanova}}, \bibinfo {author}
  {\bibfnamefont{G.}~\bibnamefont{Zamora}}, \bibinfo {author}
  {\bibfnamefont{C.~C.}\ \bibnamefont{Hilgetag}},\ and\ \bibinfo {author}
  {\bibfnamefont{J.}~\bibnamefont{Kurths}},\ }%
  \bibfield{journal}{%
  \bibinfo {journal} {Phys. Rev. Lett.}\ }%
  \textbf{\bibinfo {volume} {97}},\ \bibinfo {pages} {238103} (\bibinfo {year}
  {2006})\BibitemShut{NoStop}%
\bibitem{Galstyan07}%
  \BibitemOpen
  \bibfield{author}{%
  \bibinfo {author} {\bibfnamefont{A.}~\bibnamefont{Galstyan}}\ and\ \bibinfo
  {author} {\bibfnamefont{P.}~\bibnamefont{Cohen}},\ }%
  \bibfield{journal}{%
  \bibinfo {journal} {Phys. Rev. E}\ }%
  \textbf{\bibinfo {volume} {75}},\ \bibinfo {pages} {036109} (\bibinfo {year}
  {2007})\BibitemShut{NoStop}%
\bibitem{Buldyrev10}%
  \BibitemOpen
  \bibfield{author}{%
  \bibinfo {author} {\bibfnamefont{S.~V.}\ \bibnamefont{Buldyrev}}, \bibinfo
  {author} {\bibfnamefont{R.}~\bibnamefont{Parshani}}, \bibinfo {author}
  {\bibfnamefont{G.}~\bibnamefont{Paul}}, \bibinfo {author}
  {\bibfnamefont{H.~E.}\ \bibnamefont{Stanley}},\ and\ \bibinfo {author}
  {\bibfnamefont{S.}~\bibnamefont{Havlin}},\ }%
  \bibfield{journal}{%
  \bibinfo {journal} {Nature}\ }%
  \textbf{\bibinfo {volume} {464}},\ \bibinfo {pages} {1025} (\bibinfo {year}
  {2010})\BibitemShut{NoStop}%
\bibitem{Parshani10}%
  \BibitemOpen
  \bibfield{author}{%
  \bibinfo {author} {\bibfnamefont{R.}~\bibnamefont{Parshani}}, \bibinfo
  {author} {\bibfnamefont{S.~V.}\ \bibnamefont{Buldyrev}},\ and\ \bibinfo
  {author} {\bibfnamefont{S.}~\bibnamefont{Havlin}},\ }%
  \bibfield{journal}{%
  \bibinfo {journal} {Phys. Rev. Lett.}\ }%
  \textbf{\bibinfo {volume} {105}},\ \bibinfo {pages} {048701} (\bibinfo {year}
  {2010})\BibitemShut{NoStop}%
\bibitem{Gao12}%
  \BibitemOpen
  \bibfield{author}{%
  \bibinfo {author} {\bibfnamefont{J.~X.}\ \bibnamefont{Gao}}, \bibinfo
  {author} {\bibfnamefont{S.~V.}\ \bibnamefont{Buldyrev}}, \bibinfo {author}
  {\bibfnamefont{H.~E.}\ \bibnamefont{Stanley}},\ and\ \bibinfo {author}
  {\bibfnamefont{S.}~\bibnamefont{Havlin}},\ }%
  \bibfield{journal}{%
  \bibinfo {journal} {Nat. Phys.}\ }%
  \textbf{\bibinfo {volume} {8}},\ \bibinfo {pages} {40} (\bibinfo {year}
  {2012})\BibitemShut{NoStop}%
\bibitem{Brummitt12}%
  \BibitemOpen
  \bibfield{author}{%
  \bibinfo {author} {\bibfnamefont{C.~D.}\ \bibnamefont{Brummitt}}, \bibinfo
  {author} {\bibfnamefont{K.-M.}\ \bibnamefont{Lee}},\ and\ \bibinfo {author}
  {\bibfnamefont{K.-I.}\ \bibnamefont{Goh}},\ }%
  \bibfield{journal}{%
  \bibinfo {journal} {Phys. Rev. E}\ }%
  \textbf{\bibinfo {volume} {85}},\ \bibinfo {pages} {045102} (\bibinfo {year}
  {2012})\BibitemShut{NoStop}%
\bibitem{Cho10}%
  \BibitemOpen
  \bibfield{author}{%
  \bibinfo {author} {\bibfnamefont{W.-K.}\ \bibnamefont{Cho}}, \bibinfo
  {author} {\bibfnamefont{K.-I.}\ \bibnamefont{Goh}},\ and\ \bibinfo {author}
  {\bibfnamefont{I.-M.}\ \bibnamefont{Kim}},\ }%
  \bibfield{journal}{%
  \bibinfo {journal} {arXiv}\ }%
  \textbf{\bibinfo {volume} {1010.4971}} (\bibinfo {year}
  {2010})\BibitemShut{NoStop}%
\bibitem{Leicht09}%
  \BibitemOpen
  \bibfield{author}{%
  \bibinfo {author} {\bibfnamefont{E.~A.}\ \bibnamefont{Leicht}}\ and\ \bibinfo
  {author} {\bibfnamefont{R.~M.}\ \bibnamefont{D'Souza}},\ }%
  \bibfield{journal}{%
  \bibinfo {journal} {arXiv}\ }%
  \textbf{\bibinfo {volume} {0907.0894}} (\bibinfo {year}
  {2009})\BibitemShut{NoStop}%
\bibitem{Allard09}%
  \BibitemOpen
  \bibfield{author}{%
  \bibinfo {author} {\bibfnamefont{A.}~\bibnamefont{Allard}}, \bibinfo {author}
  {\bibfnamefont{P.-A.}\ \bibnamefont{{No\"{e}l}}}, \bibinfo {author}
  {\bibfnamefont{L.~J.}\ \bibnamefont{Dub\'{e}}},\ and\ \bibinfo {author}
  {\bibfnamefont{B.}~\bibnamefont{Pourbohloul}},\ }%
  \bibfield{journal}{%
  \bibinfo {journal} {Phys. Rev. E}\ }%
  \textbf{\bibinfo {volume} {79}},\ \bibinfo {pages} {036113} (\bibinfo {year}
  {2009})\BibitemShut{NoStop}%
\bibitem{Brummitt10}%
  \BibitemOpen
  \bibfield{author}{%
  \bibinfo {author} {\bibfnamefont{C.~D.}\ \bibnamefont{Brummitt}}, \bibinfo
  {author} {\bibfnamefont{R.~M.}\ \bibnamefont{D'Souza}},\ and\ \bibinfo
  {author} {\bibfnamefont{E.~A.}\ \bibnamefont{Leicht}},\ }%
  \bibfield{journal}{%
  \bibinfo {journal} {arXiv}\ }%
  \textbf{\bibinfo {volume} {1010.0279}} (\bibinfo {year}
  {2010})\BibitemShut{NoStop}%
\bibitem{Mendiola12}%
  \BibitemOpen
  \bibfield{author}{%
  \bibinfo {author} {\bibfnamefont{A.}~\bibnamefont{Saumell-Mendiola}},
  \bibinfo {author} {\bibfnamefont{M.~{\relax \'A}.}\ \bibnamefont{Serrano}},\
  and\ \bibinfo {author} {\bibfnamefont{M.}~\bibnamefont{Bogu\~{n}\'{a}}},\ }%
  \bibfield{journal}{%
  \bibinfo {journal} {arXiv}\ }%
  \textbf{\bibinfo {volume} {1202.4087}} (\bibinfo {year}
  {2012})\BibitemShut{NoStop}%
\bibitem{Parshani11}%
  \BibitemOpen
  \bibfield{author}{%
  \bibinfo {author} {\bibfnamefont{R.}~\bibnamefont{Parshani}}, \bibinfo
  {author} {\bibfnamefont{C.}~\bibnamefont{Rozenblat}}, \bibinfo {author}
  {\bibfnamefont{D.}~\bibnamefont{Ietri}}, \bibinfo {author}
  {\bibfnamefont{C.}~\bibnamefont{Ducruet}},\ and\ \bibinfo {author}
  {\bibfnamefont{S.}~\bibnamefont{Havlin}},\ }%
  \bibfield{journal}{%
  \bibinfo {journal} {Europhys. Lett.}\ }%
  \textbf{\bibinfo {volume} {92}},\ \bibinfo {pages} {68002} (\bibinfo {year}
  {2011})\BibitemShut{NoStop}%
\bibitem{Hu11}%
  \BibitemOpen
  \bibfield{author}{%
  \bibinfo {author} {\bibfnamefont{Y.}~\bibnamefont{Hu}}, \bibinfo {author}
  {\bibfnamefont{B.}~\bibnamefont{Ksherim}}, \bibinfo {author}
  {\bibfnamefont{R.}~\bibnamefont{Cohen}},\ and\ \bibinfo {author}
  {\bibfnamefont{S.}~\bibnamefont{Havlin}},\ }%
  \bibfield{journal}{%
  \bibinfo {journal} {Phys. Rev. E}\ }%
  \textbf{\bibinfo {volume} {84}},\ \bibinfo {pages} {066116} (\bibinfo {year}
  {2011})\BibitemShut{NoStop}%
\bibitem{Tanizawa12}%
  \BibitemOpen
  \bibfield{author}{%
  \bibinfo {author} {\bibfnamefont{T.}~\bibnamefont{Tanizawa}}, \bibinfo
  {author} {\bibfnamefont{S.}~\bibnamefont{Havlin}},\ and\ \bibinfo {author}
  {\bibfnamefont{H.~E.}\ \bibnamefont{Stanley}},\ }%
  \bibfield{journal}{%
  \bibinfo {journal} {Phys. Rev. E}\ }%
  \textbf{\bibinfo {volume} {85}},\ \bibinfo {pages} {046109} (\bibinfo {year}
  {2012})\BibitemShut{NoStop}%
\bibitem{Bassett13}%
  \BibitemOpen
  \bibfield{author}{%
  \bibinfo {author} {\bibfnamefont{D.~S.}\ \bibnamefont{Bassett}}, \bibinfo
  {author} {\bibfnamefont{M.~A.}\ \bibnamefont{Porter}}, \bibinfo {author}
  {\bibfnamefont{N.~F.}\ \bibnamefont{Wymbs}}, \bibinfo {author}
  {\bibfnamefont{S.~T.}\ \bibnamefont{Grafton}}, \bibinfo {author}
  {\bibfnamefont{J.~M.}\ \bibnamefont{Carlson}},\ and\ \bibinfo {author}
  {\bibfnamefont{P.~J.}\ \bibnamefont{Mucha}},\ }%
  \bibfield{journal}{%
  \bibinfo {journal} {Chaos}\ }%
  \textbf{\bibinfo {volume} {23}},\ \bibinfo {pages} {013142} (\bibinfo {year}
  {2013})\BibitemShut{NoStop}%
\bibitem{Newman10}%
  \BibitemOpen
  \bibfield{author}{%
  \bibinfo {author} {\bibfnamefont{M.~E.~J.}\ \bibnamefont{Newman}},\ }%
  \emph{\bibinfo {title} {Networks: An Introduction}}\ (\bibinfo {publisher}
  {Oxford University Press},\ \bibinfo {address} {Oxford},\ \bibinfo {year}
  {2010})\BibitemShut{NoStop}%
\bibitem{Newman02}%
  \BibitemOpen
  \bibfield{author}{%
  \bibinfo {author} {\bibfnamefont{M.~E.~J.}\ \bibnamefont{Newman}},\ }%
  \bibfield{journal}{%
  \bibinfo {journal} {Phys. Rev. Lett.}\ }%
  \textbf{\bibinfo {volume} {89}},\ \bibinfo {pages} {208701} (\bibinfo {year}
  {2002})\BibitemShut{NoStop}%
\bibitem{Gleeson08a}%
  \BibitemOpen
  \bibfield{author}{%
  \bibinfo {author} {\bibfnamefont{J.~P.}\ \bibnamefont{Gleeson}},\ }%
  \bibfield{journal}{%
  \bibinfo {journal} {Phys. Rev. E}\ }%
  \textbf{\bibinfo {volume} {77}},\ \bibinfo {pages} {046117} (\bibinfo {year}
  {2008})\BibitemShut{NoStop}%
\bibitem{Lancichinetti08}%
  \BibitemOpen
  \bibfield{author}{%
  \bibinfo {author} {\bibfnamefont{A.}~\bibnamefont{Lancichinetti}}, \bibinfo
  {author} {\bibfnamefont{S.}~\bibnamefont{Fortunato}},\ and\ \bibinfo {author}
  {\bibfnamefont{F.}~\bibnamefont{Radicchi}},\ }%
  \bibfield{journal}{%
  \bibinfo {journal} {Phys. Rev. E}\ }%
  \textbf{\bibinfo {volume} {78}},\ \bibinfo {pages} {046110} (\bibinfo {year}
  {2008})\BibitemShut{NoStop}%
\bibitem{Piikk_url}%
  \BibitemOpen
  \bibinfo {author} {\bibnamefont{{{\sc Matlab} (and Octave) code for
  generating $P^{i,i'}_{k,k'}$ networks and examples are available online at
  {\tt http://www.ul.ie/sdcs/melnik}}.}}\BibitemShut{Stop}%
\bibitem{Note1}%
  \BibitemOpen
\bibfield{author}{%
    }%
  \bibinfo {note} {We define $P^{i,i'}_{k,k'}$ rigorously as follows: Choose a
  network edge uniformly at random, and label its end nodes (also uniformly at
  random) as $A$ and $B$. It follows that $P^{i,i'}_{k,k'}$ is the joint
  probability that $A$ is a degree-$k$ node in module $i$ and $B$ is a
  degree-$k'$ node in module $i'$. Therefore, $(2-\delta _{i,i'}\delta
  _{k,k'})P^{i,i'}_{k,k'}$ is the probability that a randomly chosen edge
  connects a degree-$k$ node in module $i$ and a degree-$k'$ node in module
  $i'$ (because, in this case, one no longer distinguishes between the nodes at
  the two ends of the edge).}\BibitemShut{Stop}%
\bibitem{Note2}%
  \BibitemOpen
  \bibinfo {note} {To do this, one can employ the following network rewiring
  algorithm: Choose an edge of the network uniformly at random. Denote its end
  nodes by $A$ and $B$, their corresponding modules by $i_A$ and $i_B$, and
  their degrees by $k_A$ and $k_B$. Choose another edge uniformly at random
  from the set of edges that have one end-node of degree $k_A$ in module $i_A$.
  This edge connects nodes $C$ and $D$ from modules $i_A$ and $i_D$, whose
  respective node degrees are $k_A$ and $k_D$. One now rewires the two chosen
  edges to obtain the edges $AD$ and $CB$ instead of $AB$ and $CD$. In applying
  this algorithm, we also take care to avoid creating multiple-edges and
  self-edges. This rewiring scheme does not affect the degrees or modules of
  the rewired nodes, but it randomizes connections between them. Applying this
  scheme repeatedly significantly reduces the density of triangles and thereby
  reduces the local clustering.}\BibitemShut{Stop}%
\bibitem{Newman03a}%
  \BibitemOpen
  \bibfield{author}{%
  \bibinfo {author} {\bibfnamefont{M.~E.~J.}\ \bibnamefont{Newman}},\ }%
  \bibfield{journal}{%
  \bibinfo {journal} {SIAM Rev.}\ }%
  \textbf{\bibinfo {volume} {45}},\ \bibinfo {pages} {167} (\bibinfo {year}
  {2003})\BibitemShut{NoStop}%
\bibitem{Barrat08}%
  \BibitemOpen
  \bibfield{author}{%
  \bibinfo {author} {\bibfnamefont{A.}~\bibnamefont{Barrat}}, \bibinfo {author}
  {\bibfnamefont{A.}~\bibnamefont{Vespignani}},\ and\ \bibinfo {author}
  {\bibfnamefont{M.}~\bibnamefont{Barthelemy}},\ }%
  \emph{\bibinfo {title} {Dynamical Processes on Complex Networks}}\ (\bibinfo
  {publisher} {Cambridge University Press},\ \bibinfo {address} {Cambridge,
  UK},\ \bibinfo {year} {2008})\BibitemShut{NoStop}%
\bibitem{Gleeson13}%
  \BibitemOpen
  \bibfield{author}{%
  \bibinfo {author} {\bibfnamefont{J.~P.}\ \bibnamefont{Gleeson}},\ }%
  \bibfield{journal}{%
  \bibinfo {journal} {Phys. Rev. X}\ }%
  \textbf{\bibinfo {volume} {3}},\ \bibinfo {pages} {021004} (\bibinfo {year}
  {2013})\BibitemShut{NoStop}%
\bibitem{Watts02}%
  \BibitemOpen
  \bibfield{author}{%
  \bibinfo {author} {\bibfnamefont{D.~J.}\ \bibnamefont{Watts}},\ }%
  \bibfield{journal}{%
  \bibinfo {journal} {Proc. Natl. Acad. Sci. U.S.A.}\ }%
  \textbf{\bibinfo {volume} {99}},\ \bibinfo {pages} {5766} (\bibinfo {year}
  {2002})\BibitemShut{NoStop}%
\bibitem{Grassberger83}%
  \BibitemOpen
  \bibfield{author}{%
  \bibinfo {author} {\bibfnamefont{P.}~\bibnamefont{Grassberger}},\ }%
  \bibfield{journal}{%
  \bibinfo {journal} {Math. Biosci.}\ }%
  \textbf{\bibinfo {volume} {63}},\ \bibinfo {pages} {157} (\bibinfo {year}
  {1983})\BibitemShut{NoStop}%
\bibitem{Centola07a}%
  \BibitemOpen
  \bibfield{author}{%
  \bibinfo {author} {\bibfnamefont{D.}~\bibnamefont{Centola}}, \bibinfo
  {author} {\bibfnamefont{V.~M.}\ \bibnamefont{Eguiluz}},\ and\ \bibinfo
  {author} {\bibfnamefont{M.~W.}\ \bibnamefont{Macy}},\ }%
  \bibfield{journal}{%
  \bibinfo {journal} {Physica A}\ }%
  \textbf{\bibinfo {volume} {374}},\ \bibinfo {pages} {449} (\bibinfo {year}
  {2007})\BibitemShut{NoStop}%
\bibitem{Centola07b}%
  \BibitemOpen
  \bibfield{author}{%
  \bibinfo {author} {\bibfnamefont{D.}~\bibnamefont{Centola}}\ and\ \bibinfo
  {author} {\bibfnamefont{M.}~\bibnamefont{Macy}},\ }%
  \bibfield{journal}{%
  \bibinfo {journal} {Am. J. Sociol.}\ }%
  \textbf{\bibinfo {volume} {113}},\ \bibinfo {pages} {702} (\bibinfo {year}
  {2007})\BibitemShut{NoStop}%
\bibitem{FB-equal}%
  \BibitemOpen
  \bibfield{author}{%
  \bibinfo {author} {\bibfnamefont{E.}~\bibnamefont{Bakshy}},\ }%
  \enquote{\bibinfo {title} {Showing support for marriage equality on
  {F}acebook},}\  (\bibinfo {year} {2013}),\ \bibinfo {note}
  {\url{https://www.facebook.com/notes/facebook-data-science/showing-support-f%
or-marriage-equality-on-facebook/10151430548593859}}\BibitemShut{NoStop}%
\bibitem{FB-equal2}%
  \BibitemOpen
  \bibfield{author}{%
  \bibinfo {author} {\bibfnamefont{B.}~\bibnamefont{State}},\ }%
  \enquote{\bibinfo {title} {The unequal adoption of equal signs},}\  (\bibinfo
  {year} {2013}),\ \bibinfo {note}
  {\url{https://www.facebook.com/notes/facebook-data-science/the-unequal-adopt%
ion-of-equal-signs/10151927935438859}}\BibitemShut{NoStop}%
\bibitem{Gleeson07a}%
  \BibitemOpen
  \bibfield{author}{%
  \bibinfo {author} {\bibfnamefont{J.~P.}\ \bibnamefont{Gleeson}}\ and\
  \bibinfo {author} {\bibfnamefont{D.~J.}\ \bibnamefont{Cahalane}},\ }%
  \bibfield{journal}{%
  \bibinfo {journal} {Phys. Rev. E}\ }%
  \textbf{\bibinfo {volume} {75}},\ \bibinfo {pages} {056103} (\bibinfo {year}
  {2007})\BibitemShut{NoStop}%
\bibitem{Melnik13}%
  \BibitemOpen
  \bibfield{author}{%
  \bibinfo {author} {\bibfnamefont{S.}~\bibnamefont{Melnik}}, \bibinfo {author}
  {\bibfnamefont{J.~A.}\ \bibnamefont{Ward}}, \bibinfo {author}
  {\bibfnamefont{J.~P.}\ \bibnamefont{Gleeson}},\ and\ \bibinfo {author}
  {\bibfnamefont{M.~A.}\ \bibnamefont{Porter}},\ }%
  \bibfield{journal}{%
  \bibinfo {journal} {Chaos}\ }%
  \textbf{\bibinfo {volume} {23}},\ \bibinfo {pages} {013124} (\bibinfo {year}
  {2013})\BibitemShut{NoStop}%
\bibitem{Gleeson09a}%
  \BibitemOpen
  \bibfield{author}{%
  \bibinfo {author} {\bibfnamefont{J.~P.}\ \bibnamefont{Gleeson}},\ }%
  \bibfield{journal}{%
  \bibinfo {journal} {Phys. Rev. E}\ }%
  \textbf{\bibinfo {volume} {80}},\ \bibinfo {pages} {036107} (\bibinfo {year}
  {2009})\BibitemShut{NoStop}%
\bibitem{Gleeson09b}%
  \BibitemOpen
  \bibfield{author}{%
  \bibinfo {author} {\bibfnamefont{J.~P.}\ \bibnamefont{Gleeson}}\ and\
  \bibinfo {author} {\bibfnamefont{S.}~\bibnamefont{Melnik}},\ }%
  \bibfield{journal}{%
  \bibinfo {journal} {Phys. Rev. E}\ }%
  \textbf{\bibinfo {volume} {80}},\ \bibinfo {pages} {046121} (\bibinfo {year}
  {2009})\BibitemShut{NoStop}%
\bibitem{Hackett11}%
  \BibitemOpen
  \bibfield{author}{%
  \bibinfo {author} {\bibfnamefont{A.}~\bibnamefont{Hackett}}, \bibinfo
  {author} {\bibfnamefont{S.}~\bibnamefont{Melnik}},\ and\ \bibinfo {author}
  {\bibfnamefont{J.~P.}\ \bibnamefont{Gleeson}},\ }%
  \bibfield{journal}{%
  \bibinfo {journal} {Phys. Rev. E}\ }%
  \textbf{\bibinfo {volume} {83}},\ \bibinfo {pages} {056107} (\bibinfo {year}
  {2011})\BibitemShut{NoStop}%
\bibitem{Gleeson10}%
  \BibitemOpen
  \bibfield{author}{%
  \bibinfo {author} {\bibfnamefont{J.~P.}\ \bibnamefont{Gleeson}}, \bibinfo
  {author} {\bibfnamefont{S.}~\bibnamefont{Melnik}},\ and\ \bibinfo {author}
  {\bibfnamefont{A.}~\bibnamefont{Hackett}},\ }%
  \bibfield{journal}{%
  \bibinfo {journal} {Phys. Rev. E}\ }%
  \textbf{\bibinfo {volume} {81}},\ \bibinfo {pages} {066114} (\bibinfo {year}
  {2010})\BibitemShut{NoStop}%
\bibitem{Gleeson08b}%
  \BibitemOpen
  \bibfield{author}{%
  \bibinfo {author} {\bibfnamefont{J.~P.}\ \bibnamefont{Gleeson}},\ }%
  \bibfield{journal}{%
  \bibinfo {journal} {Phys. Rev. E}\ }%
  \textbf{\bibinfo {volume} {77}},\ \bibinfo {pages} {057101} (\bibinfo {year}
  {2008})\BibitemShut{NoStop}%
\bibitem{Yagan12}%
  \BibitemOpen
  \bibfield{author}{%
  \bibinfo {author} {\bibfnamefont{O.}~\bibnamefont{Ya\u{g}an}}\ and\ \bibinfo
  {author} {\bibfnamefont{V.}~\bibnamefont{Gligor}},\ }%
  \bibfield{journal}{%
  \bibinfo {journal} {Phys. Rev. E}\ }%
  \textbf{\bibinfo {volume} {86}},\ \bibinfo {pages} {036103} (\bibinfo {year}
  {2012})\BibitemShut{NoStop}%
\bibitem{Payne11}%
  \BibitemOpen
  \bibfield{author}{%
  \bibinfo {author} {\bibfnamefont{J.~L.}\ \bibnamefont{Payne}}, \bibinfo
  {author} {\bibfnamefont{K.~D.}\ \bibnamefont{Harris}},\ and\ \bibinfo
  {author} {\bibfnamefont{P.~S.}\ \bibnamefont{Dodds}},\ }%
  \bibfield{journal}{%
  \bibinfo {journal} {Phys. Rev. E}\ }%
  \textbf{\bibinfo {volume} {84}},\ \bibinfo {pages} {016110} (\bibinfo {year}
  {2011})\BibitemShut{NoStop}%
\bibitem{Ikeda10}%
  \BibitemOpen
  \bibfield{author}{%
  \bibinfo {author} {\bibfnamefont{Y.}~\bibnamefont{Ikeda}}, \bibinfo {author}
  {\bibfnamefont{T.}~\bibnamefont{Hasegawa}},\ and\ \bibinfo {author}
  {\bibfnamefont{K.}~\bibnamefont{Nemoto}},\ }%
  \bibfield{journal}{%
  \bibinfo {journal} {J. Phys.: Conf. Ser.}\ }%
  \textbf{\bibinfo {volume} {221}},\ \bibinfo {pages} {012005} (\bibinfo {year}
  {2010})\BibitemShut{NoStop}%
\bibitem{Newman03c}%
  \BibitemOpen
  \bibfield{author}{%
  \bibinfo {author} {\bibfnamefont{M.~E.~J.}\ \bibnamefont{Newman}},\ }%
  \bibfield{journal}{%
  \bibinfo {journal} {Phys. Rev. E}\ }%
  \textbf{\bibinfo {volume} {67}},\ \bibinfo {pages} {026126} (\bibinfo {year}
  {2003})\BibitemShut{NoStop}%
\bibitem{Vazquez03}%
  \BibitemOpen
  \bibfield{author}{%
  \bibinfo {author} {\bibfnamefont{A.}~\bibnamefont{V\'{a}zquez}}\ and\
  \bibinfo {author} {\bibfnamefont{Y.}~\bibnamefont{Moreno}},\ }%
  \bibfield{journal}{%
  \bibinfo {journal} {Phys. Rev. E}\ }%
  \textbf{\bibinfo {volume} {67}},\ \bibinfo {pages} {015101(R)} (\bibinfo
  {year} {2003})\BibitemShut{NoStop}%
\bibitem{Newman01e}%
  \BibitemOpen
  \bibfield{author}{%
  \bibinfo {author} {\bibfnamefont{M.~E.~J.}\ \bibnamefont{Newman}}\ and\
  \bibinfo {author} {\bibfnamefont{R.~M.}\ \bibnamefont{Ziff}},\ }%
  \bibfield{journal}{%
  \bibinfo {journal} {Phys. Rev. E}\ }%
  \textbf{\bibinfo {volume} {64}},\ \bibinfo {pages} {016706} (\bibinfo {year}
  {2001})\BibitemShut{NoStop}%
\bibitem{Note3}%
  \BibitemOpen
  \bibinfo {note} {This is a compact form of a matrix of the type that we
  showed in Fig.~\ref {fig1_Piikk}. We omit rows and columns that contain only
  $0$ entries.}\BibitemShut{Stop}%
\bibitem{condon2001}%
  \BibitemOpen
  \bibfield{author}{%
  \bibinfo {author} {\bibfnamefont{A.}~\bibnamefont{Condon}}\ and\ \bibinfo
  {author} {\bibfnamefont{R.~M.}\ \bibnamefont{Karp}},\ }%
  \bibfield{journal}{%
  \bibinfo {journal} {Random Structures \& Algorithms}\ }%
  \textbf{\bibinfo {volume} {18}},\ \bibinfo {pages} {116} (\bibinfo {year}
  {2001})\BibitemShut{NoStop}%
\bibitem{Nematzadeh14}%
  \BibitemOpen
  \bibfield{author}{%
  \bibinfo {author} {\bibfnamefont{A.}~\bibnamefont{Nematzadeh}}, \bibinfo
  {author} {\bibfnamefont{E.}~\bibnamefont{Ferrara}}, \bibinfo {author}
  {\bibfnamefont{A.}~\bibnamefont{Flammini}},\ and\ \bibinfo {author}
  {\bibfnamefont{Y.-Y.}\ \bibnamefont{Ahn}},\ }%
  \bibfield{journal}{%
  \bibinfo {journal} {arXiv}\ }%
  \textbf{\bibinfo {volume} {1401.1257v1}} (\bibinfo {year}
  {2014})\BibitemShut{NoStop}%
\bibitem{Traud08}%
  \BibitemOpen
  \bibfield{author}{%
  \bibinfo {author} {\bibfnamefont{A.~L.}\ \bibnamefont{Traud}}, \bibinfo
  {author} {\bibfnamefont{E.~D.}\ \bibnamefont{Kelsic}}, \bibinfo {author}
  {\bibfnamefont{P.~J.}\ \bibnamefont{Mucha}},\ and\ \bibinfo {author}
  {\bibfnamefont{M.~A.}\ \bibnamefont{Porter}},\ }%
  \bibfield{journal}{%
  \bibinfo {journal} {SIAM Rev.}\ }%
  \textbf{\bibinfo {volume} {53}},\ \bibinfo {pages} {526} (\bibinfo {year}
  {2011})\BibitemShut{NoStop}%
\bibitem{Traud12}%
  \BibitemOpen
  \bibfield{author}{%
  \bibinfo {author} {\bibfnamefont{A.~L.}\ \bibnamefont{Traud}}, \bibinfo
  {author} {\bibfnamefont{P.~J.}\ \bibnamefont{Mucha}},\ and\ \bibinfo {author}
  {\bibfnamefont{M.~A.}\ \bibnamefont{Porter}},\ }%
  \bibfield{journal}{%
  \bibinfo {journal} {Physica A}\ }%
  \textbf{\bibinfo {volume} {391}},\ \bibinfo {pages} {4165} (\bibinfo {year}
  {2012})\BibitemShut{NoStop}%
\bibitem{Colizza05}%
  \BibitemOpen
  \bibfield{author}{%
  \bibinfo {author} {\bibfnamefont{V.}~\bibnamefont{Colizza}}, \bibinfo
  {author} {\bibfnamefont{A.}~\bibnamefont{Flammini}}, \bibinfo {author}
  {\bibfnamefont{A.}~\bibnamefont{Maritan}},\ and\ \bibinfo {author}
  {\bibfnamefont{A.}~\bibnamefont{Vespignani}},\ }%
  \bibfield{journal}{%
  \bibinfo {journal} {Physica A}\ }%
  \textbf{\bibinfo {volume} {352}},\ \bibinfo {pages} {1} (\bibinfo {year}
  {2005})\BibitemShut{NoStop}%
\bibitem{Colizza06}%
  \BibitemOpen
  \bibfield{author}{%
  \bibinfo {author} {\bibfnamefont{V.}~\bibnamefont{Colizza}}, \bibinfo
  {author} {\bibfnamefont{A.}~\bibnamefont{Flammini}}, \bibinfo {author}
  {\bibfnamefont{M.~A.}\ \bibnamefont{Serrano}},\ and\ \bibinfo {author}
  {\bibfnamefont{A.}~\bibnamefont{Vespignani}},\ }%
  \bibfield{journal}{%
  \bibinfo {journal} {Nat. Phys.}\ }%
  \textbf{\bibinfo {volume} {2}},\ \bibinfo {pages} {110} (\bibinfo {year}
  {2006})\BibitemShut{NoStop}%
\bibitem{DIP_url_both}%
  \BibitemOpen
  \bibinfo {author} {\bibnamefont{{Protein interaction network of the yeast
  {\emph{Saccharomyces cerevisae}} extracted with different experimental
  techniques and collected at the {D}atabase of {I}nteracting {P}roteins, {\tt
  http://dip.doe-mbi.ucla.edu/;
  http://sites.google.com/site/cxnets/DIP.dat}}}}\BibitemShut{NoStop}%
\bibitem{Note4}%
  \BibitemOpen
\bibfield{author}{%
    }%
  \bibinfo {note} {Admittedly, this is not the most realistic example of
  interacting networks, but we use it because it illustrates our conceptual
  point. One can make similar comparisons using more complicated algorithms to
  conjoin the Caltech and protein-interaction networks. For example, the
  conjoining edges could depend on node degrees, modular structure, or other
  properties.}\BibitemShut{Stop}%
\bibitem{Note5}%
  \BibitemOpen
  \bibinfo {note} {For $P^{i,i'}_{k,k'}$ or $E^{i,i'}$ theory to work, we need
  to conjoin networks using sufficiently many edges to avoid finite-size
  effects, which are not captured by the theories.}\BibitemShut{Stop}%
\bibitem{Watts98}%
  \BibitemOpen
  \bibfield{author}{%
  \bibinfo {author} {\bibfnamefont{D.~J.}\ \bibnamefont{Watts}}\ and\ \bibinfo
  {author} {\bibfnamefont{S.~H.}\ \bibnamefont{Strogatz}},\ }%
  \bibfield{journal}{%
  \bibinfo {journal} {Nature (London)}\ }%
  \textbf{\bibinfo {volume} {393}},\ \bibinfo {pages} {440} (\bibinfo {year}
  {1998})\BibitemShut{NoStop}%
\bibitem{Melnik11}%
  \BibitemOpen
  \bibfield{author}{%
  \bibinfo {author} {\bibfnamefont{S.}~\bibnamefont{Melnik}}, \bibinfo {author}
  {\bibfnamefont{A.}~\bibnamefont{Hackett}}, \bibinfo {author}
  {\bibfnamefont{M.~A.}\ \bibnamefont{Porter}}, \bibinfo {author}
  {\bibfnamefont{P.~J.}\ \bibnamefont{Mucha}},\ and\ \bibinfo {author}
  {\bibfnamefont{J.~P.}\ \bibnamefont{Gleeson}},\ }%
  \bibfield{journal}{%
  \bibinfo {journal} {Phys. Rev. E}\ }%
  \textbf{\bibinfo {volume} {83}},\ \bibinfo {pages} {036112} (\bibinfo {year}
  {2011})\BibitemShut{NoStop}%
\bibitem{Gleeson12}%
  \BibitemOpen
  \bibfield{author}{%
  \bibinfo {author} {\bibfnamefont{J.~P.}\ \bibnamefont{Gleeson}}, \bibinfo
  {author} {\bibfnamefont{S.}~\bibnamefont{Melnik}}, \bibinfo {author}
  {\bibfnamefont{J.~A.}\ \bibnamefont{Ward}}, \bibinfo {author}
  {\bibfnamefont{M.~A.}\ \bibnamefont{Porter}},\ and\ \bibinfo {author}
  {\bibfnamefont{P.~J.}\ \bibnamefont{Mucha}},\ }%
  \bibfield{journal}{%
  \bibinfo {journal} {Phys. Rev. E}\ }%
  \textbf{\bibinfo {volume} {85}},\ \bibinfo {pages} {026106} (\bibinfo {year}
  {2012})\BibitemShut{NoStop}%
\end{thebibliography}%

\end{document}